\documentclass[twocolumn,showpacs,pre,floatfix]{revtex4}
\usepackage{amsfonts}
\usepackage{amssymb}
\usepackage{amsmath}
\usepackage{graphicx}

\usepackage[latin1]{inputenc}

\newcommand{\ic}{\mathrm{i}}
\newcommand{\ud}{\mathrm{d}}
\newcommand{\re}{\mathrm{ Re\ }}
\newcommand{\im}{\mathrm{ Im\ }}

\newcommand{\zZ}{\mathbb{Z}}

\begin{document}

\title{Trace formula for dielectric cavities : I. General properties}
\author{{\it E. Bogomolny,  R. Dubertrand, and  \fbox{C. Schmit}}}
\affiliation{\it Universit\'e Paris-Sud, CNRS, UMR 8626,  Laboratoire de Physique Th\'eorique et Mod\`eles Statistiques,  91405 Orsay Cedex, France}
\email{remy.dubertrand@lptms.u-psud.fr}

\date{\today}

\begin{abstract}
The construction of the trace formula for open dielectric cavities is
examined in detail. Using the Krein formula it is shown that the sum over
cavity resonances can be written as a sum over classical periodic orbits for
the motion inside the cavity. The contribution of each periodic orbit  is
the product of the two factors. The first is the same as in the standard
trace formula and the second is connected with  the product of reflection
coefficients for all  points of reflection with the cavity boundary. Two
asymptotic terms of the smooth resonance counting function related with the
area and the perimeter of a convex cavity are derived. The coefficient of the
perimeter term differs from the one for closed cavities due to unusual
high-energy asymptotics of the $\mathbf{S}$-matrix for the scattering on the
cavity. Corrections to the leading semi-classical formula are briefly
discussed.  Obtained formulas agree well with numerical calculations for
circular dielectric cavities.  
\end{abstract}
\maketitle

\section{Introduction}

Only a limited number of quantum models permits exact analytical
solutions. All others require either numerical or approximate
approaches. One of the most useful approximation and widely used for
non-integrable multi-dimensional quantum systems is  the semiclassical one
based on different types of trace formulas developed in the early 70's
\cite{balian_bloch,gutzwiller_1,gutzwiller,berry}.  

A general trace formula relates two different objects. Its left-hand side
is the density of discrete energy levels, $E_m$, of a quantum system  
\begin{equation}
 d_{\mathrm{quantum}}(E)=\sum_m\delta(E-E_m)\ .
\label{delta}
\end{equation}
The right-hand side of the trace formula is the sum of two contributions
\begin{equation}
 d_{\mathrm{semi-classical}}(E)=\bar{d}(E)+ d^{(osc)}(E)\ .
 \label{classical}
\end{equation}
The first term, $\bar{d}(E)$, is the smooth part of level density given by a
series of the Weyl terms (see e.g. \cite{balian_bloch} and references
therein). By definition, $\bar{d}(E) =\mathrm{d}\bar{N}(E)/\mathrm{d}E$
where  $\bar{N}(E)$ is the mean number of levels with energies $E_n\leq E$
called also the smooth counting function.   For two-dimensional billiards
leading contributions  to $\bar{N}(E)$ are the following
\begin{equation}
 \bar{N}(E)=\frac{A k^2}{4\pi}+r\frac{Lk}{4\pi }+\mathcal{O}(1)
\label{weyl}
\end{equation}
where $k=\sqrt{E}$ is the momentum, $A$ is the area of the billiard cavity,
and $L$ is its perimeter. $r$ in this formula is related with chosen
boundary conditions. For example,  $r=\pm 1$ for, respectively, the Neumann
and the Dirichlet conditions.  

The second  term in (\ref{classical}), $d^{(osc)}(E)$, is the oscillating
(or  the fluctuating) part of the level density given in the leading order
by the sum over classical periodic orbits 
\begin{equation}
 d^{(osc)}(E) =\sum_{\mathrm{periodic\  orbits}}d_p(E)+\mathrm{c. c.}
\end{equation} 
where  $d_p(E)$ is the contribution of a given periodic orbit
\begin{equation}
 d_p(E)=c_p\mathrm{e}^{\mathrm{i}S_p(E)/\hbar}\ .
\end{equation}
Here $S_p(E)$ is the classical action along the periodic orbit (for
billiards $S_p(E)=kl_p$ with $l_p$ being the length of the periodic orbit)
and the prefactor $c_p$ can be computed from classical characteristics of
the given periodic orbit (see below (\ref{dE_chao}) and (\ref{dE_int})). The
trace formula is the statement that 
\begin{equation}
d_{\mathrm{quantum}}(E) \stackrel{\hbar\to
  0}{\longrightarrow}d_{\mathrm{semi-classical}}(E)\ . 
\label{tp}
\end{equation}
In an overwhelming majority of cases the trace formulas were applied to
closed quantum systems with discrete energy levels. Though it is known (see
e.g. \cite{gaspard}) that the trace formula and its modifications can also
be applied for open quantum systems,  only a small number of examples has
been considered so far \cite{cvitanovic}. 

This is the first of a series of papers whose purpose is to demonstrate the
usefulness of the application of trace formulas to open quantum systems, in
particular, to dielectric cavities where a bounded domain is filled with a
dielectric material with refractive index  $n>1$ while the exterior is a
media  with refractive index $1$. The increasing interest to the
investigation of such models to a large extend  is related  with their
potential importance as resonators in micro-lasers \cite{nockel,stone} and
the experimental possibilities of observing geometrical characteristics of
the cavity from their lasing spectra \cite{melanie,these_Melanie}. The
construction  of trace formulas for dielectric cavities has been briefly
discussed in \cite{these_Melanie}. We also mention that
for open chaotic resonators a variant of a trace formula has been developed
in \cite{trace_openb}.

Open systems, in general, have no discrete spectrum. So it is not
immediately clear what should be  the left-hand side (the spectral part) of
the trace formula. In Section~\ref{general} we discuss the Krein formula
\cite{krein_1} which is the main theoretical tool for open systems. This
formula relates the excess density of states for an open cavity with the
derivative over energy of the determinant of the $\mathbf{S}$-matrix for the
scattering on the cavity. From the Krein formula it follows that the
spectral part of the trace formula consists of the sum over all complex
poles  of the $\mathbf{S}$-matrix widely called  resonances or
quasi-stationary states. 

The right-hand side of the trace formula should include a smooth part and a
sum over periodic orbits (cf. (\ref{classical})).  The general formalism
like the multiple scattering approach, used by Balian and Bloch in their
construction of the trace formula for closed cavities \cite{balian_bloch},
is not yet fully developed for dielectric cavities. The main difference of
dielectric cavities from closed ones is that for the later one knows
boundary conditions on the cavity boundary but for the former boundary
conditions are fixed only at infinity.  Due to this fact the resulting
integral equation (\ref{integral_equation}) includes the integration over
the cavity volume and not over its boundary as for the closed case which
complicates the general construction. In Section~\ref{properties} the form
of leading  periodic orbit contributions is fixed from  physical
considerations. 

To get more insight to this problem, in Section~\ref{circle} a simple
example of integrable circular dielectric cavity is considered in
detail. Despite the calculations are straightforward, they permit us to
calculate in Section~\ref{weyl_term} the smooth part of the counting
resonance function and in Section~\ref{periodic} the periodic orbit
contributions. In Section~\ref{goos} we consider main corrections to the
asymptotic results and, in particular, demonstrate how the Goos-Hänchen
shift \cite{g_h} manifests in the trace formula. All obtained formulas agree
well with numerical calculations. 

Though  these results were obtained, strictly speaking, only for circular
dielectric cavities, all main steps leading to them are quite general and we
conjecture that they remain valid for dielectric cavities of arbitrary
shape.  The detailed comparison of the derived  trace formulas with
numerical calculations for cavities of different shapes is postponed for a
future publication \cite{part_II}.  

\section{Krein formula} \label{general}

Throughout this paper we  focus on a planar two-dimensional domain
$\mathcal{D}$. The wave equations widely used for such dielectric cavities
are the following (see e.g. \cite{jackson}) 
\begin{eqnarray}
 (\Delta + n^2 E ) \Psi(\vec{x}\,)&=&0 \mbox{ when } \vec{x}\in\mathcal{D}\
 ,\nonumber\\ 
(\Delta +  E ) \Psi(\vec{x}\,)&=&0 \mbox{ when } \vec{x} \notin\mathcal{D} \ .
\label{eqopenbill}
\end{eqnarray}
In electromagnetism we rather consider the wavenumber $k=\sqrt{E}$, which is
related to the frequency $\omega$ of the wave through $k=\omega/c$, c being
the speed of light in the vacuum.
For simplicity we shall suppose that the wave function, $\Psi(\vec{x}\,)$,
and its normal derivative are continuous along the boundary of
$\mathcal{D}$. In electrodynamics these boundary conditions correspond to
transverse magnetic  (TM) modes inside an infinite dielectric cylinder with
cross section  $\mathcal{D}$ \cite{jackson}. Transverse electric (TE)  modes
may be treated similarly and will be considered elsewhere.  

We stress that Eqs.~(\ref{eqopenbill}) describe correctly electromagnetic
fields only for an infinite dielectric cylinder. For a real 3-dimensional
cavity with a small height, $h$, they have to be modified. The authors are
not aware of the full 3-dimensional treatment of such problems. The usual
approach consists to consider the refractive index, $n$, in these equations
not as a constant but as an  effective refraction index, $n=n_{eff}(kh)$,
for the motion inside a 2 dimensional dielectric slab (see e.g. \cite{nous}
and references therein) but errors of a such approximation are not known at
present.  

The possibility of using such cavities as resonators is related with the
phenomenon of the total internal reflection. It is known (see
e.g. \cite{jackson}) that if the TM plane wave inside the cavity is
reflected from a straight boundary, the (Fresnel) coefficient of reflection
has the following form  
\begin{equation}
R(\theta)=
\left\{ \begin{array}{cc}\dfrac{n\cos \theta-\sqrt{1-n^2\sin^2
        \theta}}{n\cos \theta+\sqrt{1-n^2\sin^2 \theta}} , &\textrm{ when
    }|\theta|<\theta_c\\ 
 \dfrac{n\cos \theta-\ic \sqrt{n^2\sin^2 \theta-1}}{n\cos \theta+\ic
   \sqrt{n^2\sin^2 \theta-1}}, &\textrm{ when }|\theta|>\theta_c
                  \end{array}\right .\ .
\label{reflection}
\end{equation}
Here $\theta$ is the angle between the direction of the incoming wave and
the normal to the boundary and the critical angle 
\begin{equation}
 \theta_c=\arcsin 1/n\ .
 \label{critical}
 \end{equation}
 When $|\theta|>\theta_c$ $|R(\theta)|=1$ and the wave is completely
 reflected from the boundary which may lead to formation of long-lived
 states. We mention that this expression becomes less efficient for curved
 boundary close to the critical angle (see e.g. \cite{felsen_marcuvitz}). 

Eq.~(\ref{eqopenbill}) admits only continuous spectrum and its properly
renormalized eigenvalue density, $d(E)$, has to be a smooth function of
energy, contrary to closed systems where the level density  is a sum of
delta's peaks (cf. (\ref{delta})). It is convenient to rewrite
(\ref{eqopenbill}) as the Schr\"odinger  equation with a potential  
\begin{equation}
 (\Delta +  E ) \Psi(\vec{x}\,)= -\lambda V(\vec{x}\,)\Psi(\vec{x}\,)
\label{convenient}
\end{equation}
where the coupling constant
\begin{equation}
 \lambda=E(n^2-1)
\label{lambda}
\end{equation}
and the potential $V(\vec{x}\,)$ is non-zero only inside the cavity 
\begin{equation}
 V(\vec{x}\,)=\left \{\begin{array}{cc} 1&\mbox{ when } \vec{x}\in\mathcal{D}\\ 
0& \mbox{ when } \vec{x} \notin\mathcal{D} \end{array}\ .\right .
\label{potential}
\end{equation}
Except of unusual dependence of the coupling constant on the energy this
equation describes the motion in a finite-range potential so  standard
methods (see e.g. \cite{s_matrix}) can be applied to analyze it. In
particular, its solution corresponding to the continuous spectrum is defined
by the following  integral equation (see also \cite{rawlings}) 
\begin{equation}
 \Psi(\vec{x}\,)=\mathrm{e}^{\ic \vec{k}\vec{x}}-\lambda
 \int_{\mathcal{D}}G_0(k|\vec{x}-\vec{y}\,|)\Psi(\vec{y}\,)\mathrm{d}\vec{y} 
 \label{integral_equation}
\end{equation}
where $\vec{k}$ is the momentum vector of the incoming wave with coordinates 
$k(\cos\, \theta,\sin\, \theta)$, $G_0(k|\vec{x}-\vec{y}\, |)$ is the free
Green function of the left-hand side of (\ref{convenient}) 
\begin{equation}
 G_0(z)=\frac{1}{4\ic}H^{(1)}_0(z),
\end{equation}
and $H^{(1)}_0(z)$ is the Hankel function of the first order.  

This equation may serve as a starting point of the multiple scattering
method similar to the one discussed in \cite{balian_bloch}. The important
difference between these two cases is that the integration in
(\ref{integral_equation}) is performed over the whole cavity volume which
complicates the iteration procedure. Nevertheless,
Eq.~(\ref{integral_equation}) has the standard form of the Fredholm equation
and its solutions are  well defined similar to the usual case when  coupling
constant $\lambda$ is independent of energy. In particular,  from
(\ref{integral_equation}) it follows that the $\mathbf{S}$-matrix for the
scattering on the cavity has the form 
\begin{equation}
 \mathbf{S}(\theta,\theta^{\prime})=\delta(\theta-\theta^{\prime})+\frac{\ic
   \lambda}{4\pi}f(\theta,\theta^{\prime}) 
\end{equation}
where $\theta$ and $\theta^{\prime}$ are the angles determining the
directions of, respectively, incoming and outcoming waves and  
\begin{equation}
 f(\theta,\theta^{\prime})=\int_{\mathcal{D}}\mathrm{e}^{-\ic
   \vec{k^{\prime}}\vec{y}}\Psi(\vec{y}\,)\mathrm{d}\vec{y}\ . 
\end{equation}
Here $\vec{k^{\prime}}$ is the momentum of the outgoing wave with
coordinates $k(\cos \theta^{\prime},\sin \theta^{\prime})$. 

The importance of the $\mathbf{S}$-matrix lies in the fact  that 
the excess density of states for open systems can conveniently be written
through it by using the Krein formula (see \cite{krein_1} and
\cite{lifshitz,birman_1}).  This formula relates the density of states of
two operators: the first with a short-range potential, $d(E)$, and the
second without it, $d_0(E)$. In physical terms it reads 
\begin{equation}
  d(E)-d_0(E)=\frac{1}{2\pi\ic}\frac{\partial }{\partial E}\ln \ \mathrm{
    det }\ \mathbf{S}(E) 
\label{fkrein}
\end{equation}
where $\mathrm{det}\ \mathbf{S(E)}$ is the determinant of the
$\mathbf{S}$-matrix for the scattering on the potential. For clarity in
Appendix~\ref{krein1dim} a physical  'derivation' of this formula is given
in the simplest case of one-dimensional systems.  

It is also known (see e.g. \cite{s_matrix}, sect. 12, and
\cite{zworski_3,poisson}) that function det $ \mathbf{S}(E)$ for the
scattering on a finite-range potential  is a meromorphic function in the
complex $k\equiv \sqrt{E}$ plane with (in even-dimensional spaces) a  cut
along the negative $k$-axis 
\begin{equation}
 \mathrm{ det }\ \mathbf{S}(E)=\mathrm{e}^{\ic f(k)}\prod_m\frac{k-k_m^*}{k-k_m}
\label{meromorphic}
\end{equation}
where $k_m$ with Im $k_m<0$ denote the positions of the poles of the
$\mathbf{S}$-matrix on the second sheet of energy surface and the product is
taken over all such poles. (Terms which ensure the convergence of this
infinite product are not written explicitly).  Here $f(k)$ is a certain
function without singularities in the cut complex plane related with the
asymptotics of the $\mathbf{S}$-matrix when $E\to \infty$.  Due to the
symmetry $k\to -k$ of (\ref{eqopenbill}),  if $k_m$ is a pole, $-k_m^*$ will
also be a pole which is implicitly assumed in (\ref{meromorphic}).  

Therefore, the left-hand side of (\ref{fkrein}) can be written as the sum
over all poles of the $\mathbf{S}$-matrix 
\begin{eqnarray}
  d(E)-d_0(E)&=&g(E)-\frac{1}{2\pi k} \sum_m \im \frac{1}{k-k_m}\nonumber\\
&=&g(E)-\frac{1}{\pi}\sum_m \im \frac{1}{E-E_m}
\label{sum_poles}
\end{eqnarray}
where $E_m=k_m^2$ and $g(E)=f^{\prime}(k)/2\pi $.
In mathematical literature such formulas are known as Poisson formulas (see
\cite{poisson} and references therein). 
 
From  (\ref{sum_poles}) it is clear that poles of the $\mathbf{S}$-matrix,
especially those close to the real axis, play an important role in $d(E)$ as
they produce  peaks in the otherwise smooth background.  

The positions of these poles, called also resonances or quasi-stationary
states,  are in principle determined from Eq.~(\ref{eqopenbill}) by an
analytic continuation of the $\mathbf{S}$-matrix from the real axis to the
complex plane, in a similar way as in \cite{main_wiersig}. Another way is to
impose the outgoing radiation conditions (see e.g. \cite{s_matrix})  
\begin{equation}
  \Psi(\vec{x}\,) \propto  \mathrm{e}^{{ \ic} k |\vec{x}|}\; \mbox{ when }
  |\vec{x}| \to \infty \ . \label{emergent}
\end{equation}
These conditions select a well defined set of complex eigenvalues $E_m$
whose imaginary parts correspond to the resonance lifetime $\tau_m$, $  \im
E_m=-1/2\tau_m$. 

These arguments demonstrate that the spectral part of the trace formula (the
analog of (\ref{delta})) may be written as the  sum over all poles (=
resonances) of the $\mathbf{S}$-matrix 
\begin{eqnarray}
&&d_{\mathrm{quantum}}(E)=-\frac{1}{\pi }\sum_m \frac{\im E_m}{(E-\re
  E_m)^2+(\im E_m)^2}\nonumber\\ 
&& = -\frac{1}{2\pi k}\sum_m \frac{\im k_m}{(k-\re k_m)^2+(\im k_m)^2}\ .
\label{sum_poles_all}
\end{eqnarray}
When Im $E_m\to 0$ one recovers the usual $\delta$-function contribution.

\section{General properties of trace formulas for dielectric
  cavities}\label{properties} 

To find the right-hand side of trace  formula one has to express it through
the Green function. In this section we will only consider the real
continuous spectrum of Eq.~(\ref{eqopenbill}) and eigenfunctions associated
with it. We will not deal with functions associated with the resonances
defined by boundary condition (\ref{emergent}). A minor difference with the
usual case of  closed systems consists in the fact that the eigenfunctions
are orthogonal not with respect to the standard scalar 
product $\int \Psi_{E_1}^*(\vec{r}\,)\Psi_{E_2}(\vec{r}\,)\ud\vec{r}$ 
but to a modified one
\begin{equation}
  \int n^2(\vec{r}\,)
  \Psi_{E'}^*(\vec{r}\,)\Psi_{E''}(\vec{r}\,)\ud\vec{r}=\delta(E'-E'')
\end{equation}
where the integration is performed over the whole space. Here
$n(\vec{r}\,)=n$ for points  inside the cavity and $n(\vec{r}\,)=1$ outside
it. This relation can  easily  be checked from the main equation
(\ref{eqopenbill}).  

This modification leads to the following formal relation between the density of states and  the trace of the Green function (cf. e.g. \cite{balian_bloch})
\begin{equation}
  d(E)=-\frac{1}{\pi}\int n^2(\vec{r}\,)\im G_E(\vec{r},\vec{r}\, ) \ud\vec{r}\ 
\label{defde}
\end{equation}
where the integration is extended to the whole space. 

Exactly as for closed systems (see
e.g. \cite{balian_bloch,gutzwiller_1,gutzwiller}), the dominant contribution
to the fluctuating part of the density of states comes from classical
periodic orbits. When saddle point is considered, it is clear that the only
information that a trajectory may have about a boundary is contained in the
reflection coefficient from this boundary. Therefore, the principal
difference with the closed systems is that the contribution of a given
periodic orbit has to be multiplied by the product of reflection
coefficients (\ref{reflection}) for all bounces with the cavity boundary
(which we called the total reflection coefficient) and the later can be less
than unity.  

These simple arguments demonstrate that for dielectric cavities  the
dominant contribution of a periodic orbit to the density of states when
$k\to \infty$  has  the following form  
\begin{itemize}
\item For an isolated primitive periodic orbit $p$ repeated $r$ times
  \begin{equation}
    \label{dE_chao}
    d_p(E)=\frac{n l_p}{\pi k\ |\det(M_p^r-1)|^{1/2}} R_p^r \mathrm{e}^{\ic
      [r n k l_p- r \mu_p\pi/2]} 
  \end{equation}
where  $l_p,M_p, \mu_p$, $R_p$  are, respectively, the length, the monodromy
matrix, the Maslov index, and  the total reflexion coefficient for the
chosen primitive periodic orbit. 
\item For a primitive periodic orbit  family   
  \begin{equation}
    \label{dE_int}
  d_p(E)=\frac{n^{3/2}\mathcal{A}_p}{\pi \ \sqrt{2\pi r k l_p}} \langle
  R_p^r \rangle \mathrm{e}^{\ic [ r k l_p- r\mu_p\pi/2+\pi/4]} 
  \end{equation}
where   $\mathcal{A}_p$ is the area covered by  periodic orbit family,
$\langle R_p^r \rangle$ is the mean over family value of the total
reflection coefficient, and all other notations are as above. 
\end{itemize}

The dependence on $n$ of the prefactors in these formulas is related with
the fact that inside the cavity the momentum equals $nk$.  

Now we have all ingredients of the trace formula for dielectric cavities
except the smooth (Weyl) terms (\ref{weyl}). It is clear that its value
depends on how many $\mathbf{S}$-matrix resonances  are included in the
right-hand side of the trace formula (\ref{sum_poles_all}). Energy
eigenvalues of closed systems are real and they all have to be included. For
open systems resonance energies are complex and a natural approach is to
take into account only resonances whose imaginary part is restricted,
e.g. $-\im k_m <\gamma$. Other resonances (if any) have to be considered as
a smooth background.  This separation of poles is, to a large extent,
arbitrary which manifests in the fact that the smooth part of the resonance
density will be now  a function of $\gamma$ whose calculation is a difficult
problem.  

For 2-dimensional open chaotic billiard problems with holes there exist
strong arguments \cite{zworski_1,zworski_2} that the leading term of the
Weyl law is non-trivial 
\begin{equation}
 \bar{N}(E)\sim C(\gamma)k^{\nu}
\label{fractal}
\end{equation}
where $\nu<2$ is related with the fractal dimension of the trapped set of
classical orbits.  

For open dielectric cavities arguments leading to (\ref{fractal}) can not be
directly applied and  it appears \cite{nonnenmacher} that in this case
standard estimates (\ref{weyl}) with different constants are valid. In
general, there exist two types of resonances whose behavior is  different
when the system is somehow  "closed".  The first (sometimes called Feshbach
or internal resonances) are resonances which tends to eigenvalues of the
corresponding closed system and the second ones (called shape  or outer
resonances) are those whose imaginary part remains non-zero when a system is
closed. The both types of resonances exist for dielectric cavities (see
e.g. \cite{nous}). In all cases with convex shaped cavities considered by us
these two groups of 
resonances are  well separated and  there exists a clearly defined value of
$\gamma_{\mathrm{max}}$ such that all first type resonances have $-\im
k_m<\gamma_{\mathrm{max}}$. It does not contradict \cite{main_wiersig} where
a Fractal Weyl law was argued to be a good description of a part of the
resonance spectrum with small imaginary part. In the next section we shall
argue that  when 
all these resonances are taken into account the smooth counting function
$\bar{N}(E)$ defined as the mean number of resonances with
$\mathrm{Re}\,E_n<E\equiv k^2$ and $-\im k_m<\gamma_{\mathrm{max}}$  is
similar to  (\ref{weyl}) but with  following modifications  
\begin{equation}
 \bar{N}(E)=\frac{An^2 k^2}{4\pi}+\tilde{r}(n)\frac{L k}{4\pi }+\mathcal{O}(1) 
\label{weyl_dielc}
\end{equation}
where  $A$ and $L$ are, as above, the area and the perimeter of the cavity
but   $\tilde{r}(n)$ is 
\begin{equation}
 \tilde{r}(n)=1+\frac{n^2}{\pi}\int_{-\infty}^{\infty}
 \frac{\mathrm{d}t}{t^2+n^2}R(t)-  
 \frac{1}{\pi}\int_{-\infty}^{\infty}\frac{\mathrm{d}t}{t^2+1}R(t)
 \label{new_weyl}
\end{equation}
and 
\begin{equation}
R(t)=\dfrac{\sqrt{t^2+n^2}-\sqrt{t^2+1}}{\sqrt{t^2+n^2}+\sqrt{t^2+1}}\ .
\label{new_coeff}
\end{equation}
The function $r(n)$ can be expressed through elliptic integrals. It is
monotonic function of $n$ starting from $1$ for $n=1$ and tending to
$n(4/\pi-1)$ for large $n$.  

Finally,  the trace formula for dielectric cavities has the following form
\begin{eqnarray}
&-&\frac{1}{\pi}\sum_{-\im k_m<\gamma_{\mathrm{max}}}\frac{\im k_m}{(k-\re k_m)^2+(\im k_m)^2}\nonumber\\
&=&2k\{\bar{d}(E)+\sum_{p}[d_p(E)+d_p^*(E)]\}\ . 
\label{dk}
\end{eqnarray}
Here $d_p(E)$ for isolated periodic orbits is given by (\ref{dE_chao}) and
for periodic orbits from a family it is defined in (\ref{dE_int}). 
The factor $2k$  in the right-hand side is introduced as we found it more
convenient to work with the density of states as a function of momentum.  

To see the effect of periodic orbit it is usual to multiply the both sides
of the trace formula (\ref{dk}) by a test function, e.g. by
$\mathrm{e}^{-\mathrm{i} k nl}$, and integrate over $k$ in a certain window
$k_1<k<k_2$ which includes many resonances. The dominant contribution in the
left-hand side  
of this formula is  the sum over resonances whose momentum is restricted
$k_1<\re k<k_2$ and $0<-\im k_m<\gamma_{\mathrm{max}}$ 
\begin{equation}
\sum_m\mathrm{e}^{-\ic k_m n l }\simeq \sum_p I_p(l)
\label{fourier}
\end{equation}
where the summation is done over periodic orbits, $p$, and it is implicitly assumed that term with $p=0$ corresponds to the smooth part of the trace formula. 
Here $I_p$ denotes the integral the individual periodic orbit contribution
given by (\ref{dE_chao}) or (\ref{dE_int}) 
\begin{equation}
 I_p(l)= \int_{k_1}^{k_2}\mathrm{e}^{-\mathrm{i}knl}d_p(E)2k\mathrm{d}k
\label{beta}
\end{equation}
which is strongly peaked at the periodic orbits length. Eq.~(\ref{fourier})
is called the length density. 

Periodic orbits are not the only contributions to the trace formula. For
polygonal cavities  important contributions are given by diffractive orbits
which go through  singularities of the boundary (see e.g. \cite{pavloff} and
references therein). Usually their individual contribution is smaller than
those of periodic orbits. The careful calculation of such  corrections
requires the knowledge of the diffraction coefficient on dielectric
singularities which is not available analytically. We shall also not discuss
here creeping orbits (see e.g. \cite{creeping} and references therein)
corresponding to the external motion along cavity boundary as they are
responsible only for shape resonances and their contributions are small.  

\section{Circular cavity}\label{circle}

The circular dielectric cavity is one example of a two-dimensional  cavity
with an explicit analytical solution and it is instructive to illustrate the
above general formulas in this simple case.  

The Green function $G(\vec{r},\vec{r}^{\ \prime},E)$ for this problem has
been written  e.g. in \cite{nous}.  
From that formulas it follows that inside the circular cavity with radius
$R$ the Green function  $G(\vec{r},\vec{r},E)$ with $r<R$ has the  form 
\begin{eqnarray}
G(\vec{r},\vec{r}\,)&=&
-\frac{1}{2\pi x}\sum_{m\in \zZ} \frac{J_m^2(nkr)}{s_m(x)J_m^2(x)}\nonumber\\
 &+& \frac{1}{4\ic} \sum_{m\in\zZ} \frac{J_m(nkr)}{J_m(nx)}g_{in}(r)
\end{eqnarray}
where
\begin{equation}
g_{in}(r)=J_m(nx)H^{(1)}_m(nkr) -  J_m(nkr) H^{(1)}_m(nx)\ ,
\end{equation}
and outside it (when $r>R$) it is
\begin{eqnarray}
 G(\vec{r},\vec{r}\,) &=&
-\frac{1}{2\pi x}\sum_{m\in \zZ} \frac{H^{(1)\ 2}_m(kr)}{s_m(x)H^{(1)\
    2}_m(x)}\label{outside}\\ 
 &+& \frac{1}{4\ic} \sum_{m\in\zZ}
 \frac{H^{(1)}_m(kr)}{H^{(1)}_m(x)}g_{out}(r) \nonumber 
\end{eqnarray}
with
\begin{equation}
g_{out}(r)=H^{(1)}_m(x)J_m(kr) -  J_m(x) H^{(1)}_m(kr)\ .
\end{equation}
Here $x\equiv kR$, $J_m(z)$ (resp. $H^{(1)}_m(z)$) stands for the Bessel
function (resp. the Hankel function of the first kind), and 
\begin{equation}
  s_m(x)=n\frac{J_m^{\prime}}{J_m}(nx)-\frac{H^{(1)\prime}_m}{H_m^{(1)}}(x)\ .
  \label{Sm} 
\end{equation}
Quantities $s_m(x)$ are of special importance as the positions of
resonances, $x_m\equiv k_m R$,  are  determined by their complex  zeros (see
e.g. \cite{nous}) 
\begin{equation}
s_m(x_m)=0\ .
\label{zeros}
\end{equation}
Using the formula \cite{bateman}
\begin{eqnarray}
  \int^x A_\nu(t)B_\nu(t)t \ud t&=& \frac{x^2}{2}A_\nu'(x)B_\nu'(x)\nonumber\\
 &+&  \frac{x^2-\nu^2}{2}A_\nu(x)B_\nu(x)
\label{intbess}
\end{eqnarray}
valid for any two Bessel functions  $A_\nu(z)$ and $B_\nu(z)$ of the same
arguments and performing straightforward calculations one gets 
\begin{equation}
  d(E)-d_0(E)= \frac{R^2(n^2-1)}{2\pi x} \sum_{m=-\infty}^{\infty} \im
  \frac{1}{s_m(x)}  
\label{d_circle}
\end{equation}
where $d(E)$ is formally
\begin{equation}
 d(E)=2\pi\Big [ n^2\int_0^R +\int_R^A\Big ] \im G(\vec{r},\vec{r}\,)r \ud r
\end{equation}
and 
\begin{equation}
 d_0(E)=2\pi \int_0^A \im G_0(\vec{r},\vec{r}\,)r \ud r
\end{equation}
with the free Green function 
\begin{equation}
 G_0(\vec{r},\vec{r}^{\
   \prime})=\frac{1}{4\ic}H^{(1)}_0(k|\vec{r}-\vec{r}^{\ \prime}|)\ . 
\end{equation}
In the above formulas it is assumed that the cut-off $A\gg R$ and the limit
$A\to \infty$ is taken after the subtraction $d(E)-d_0(E)$. 

The calculation of the $\mathbf{S}$-matrix for the circular dielectric
cavity is also well known \cite{debye}  and one finds 
\begin{equation}
  \mathbf{S}_m(E)=-\frac{H^{(2)}_m(x)s_m(x)^*}{H^{(1)}_m(x)s_m(x)}
  \label{matS}
\end{equation}
with $s_m(x)$ are defined in (\ref{Sm}). 
 It is straightforward  to check that (\ref{d_circle}) can be written in the form
\begin{equation}
  \label{krein_cirlce}
  d(E)-d_0(E)=\frac{1}{2 \pi\ic} \frac{\partial\ }{\partial E} \ln \det \
  \mathbf{S}(E)   
\end{equation}
which is the Krein formula (\ref{fkrein}) for the dielectric disk.
  
\begin{figure}
\begin{minipage}{.45\linewidth}
\includegraphics[width=.9\linewidth,angle=-90]{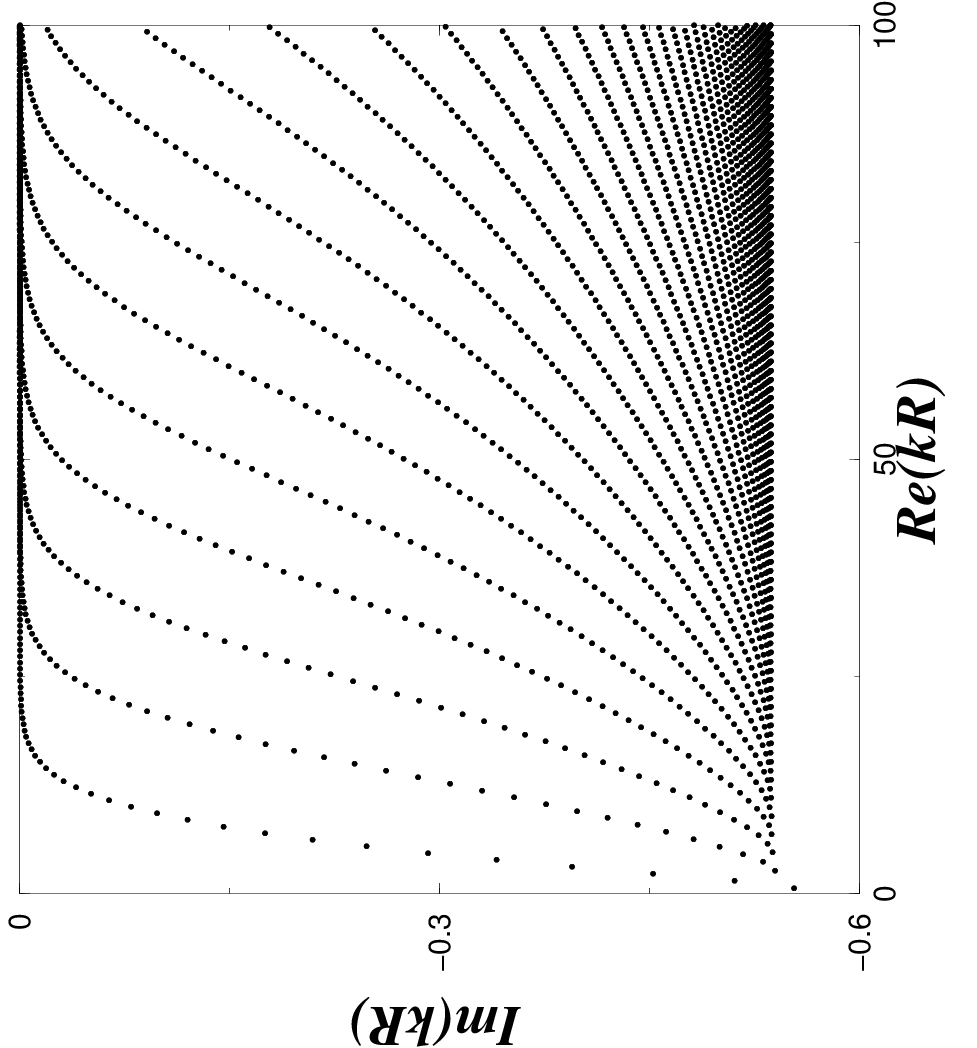} 
\begin{center}a)\end{center}
\end{minipage}\hfill
\begin{minipage}{.54\linewidth}
\includegraphics[width=.88\linewidth]{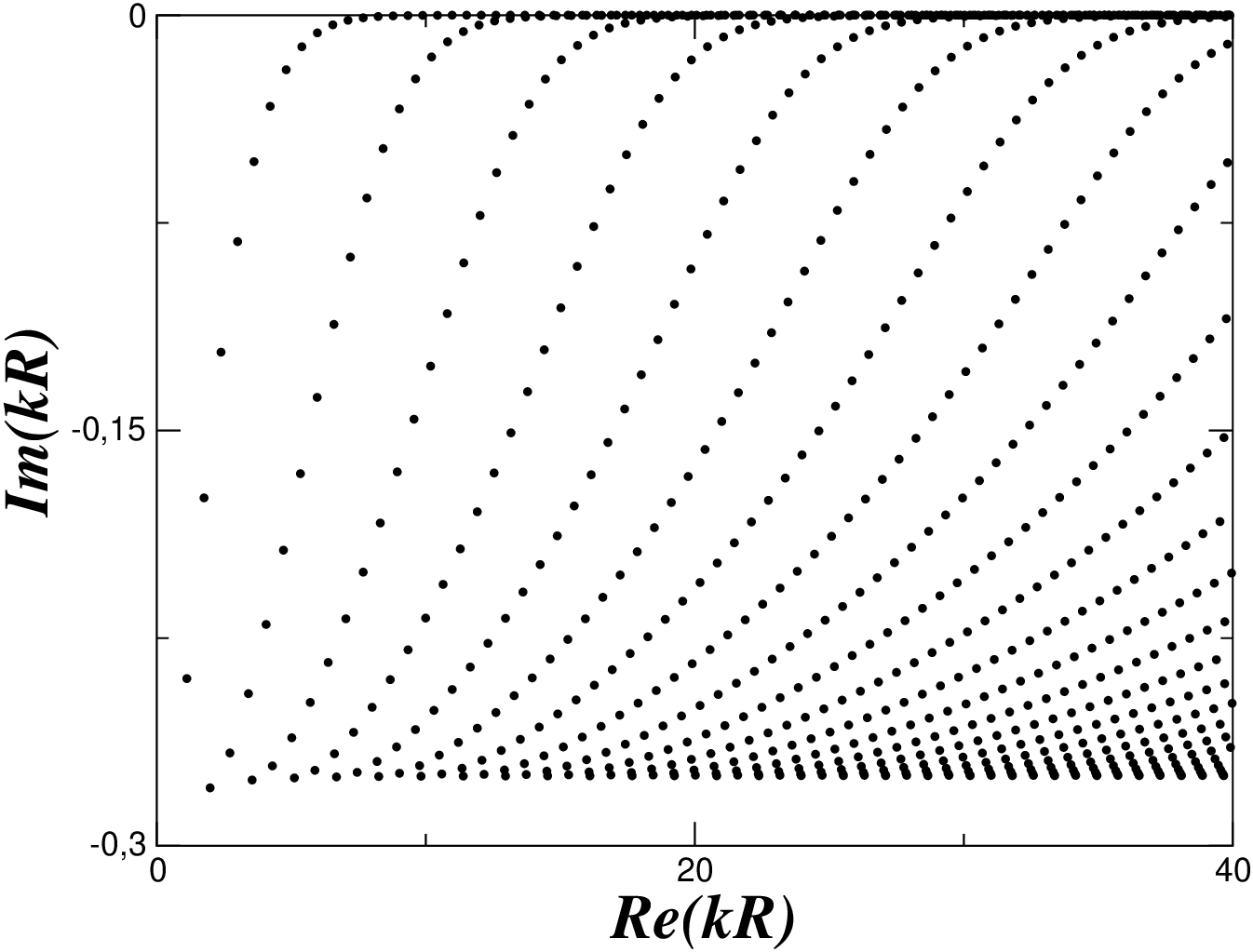}
\begin{center}b)\end{center}
\end{minipage}
\caption{a) Resonance spectrum for the dielectric circular cavity with
  $n=1.5$ and Re $kR<100$. b) The same but for $n=2$ and Re $kR<40$.  }  
\label{spectrum_disk}
\end{figure}

 In Figs.~\ref{spectrum_disk}a) and b)  we present numerically computed from
 (\ref{zeros}) positions of complex  momenta of  quasi-bound states for
 dielectric circular cavities with $n=1.5$ and $n=2$.      
Clearly, resonances in this case are organized in families corresponding to
conserved quantum numbers of radial and angular momenta (see
e.g. \cite{nous}). Let notice that for large Re $kR$  imaginary parts of all
resonances obey the inequality 
\begin{equation}
-\mathrm{Im}\ kR\leq \gamma_{\mathrm{max}}(n)
\label{max_gamma}
\end{equation} 
where 
\begin{equation}
\gamma_{\mathrm{max}}(n)=\frac{1}{2n}\ln \frac{n+1}{n-1}\ .
\label{limit}
\end{equation}
$\gamma_{\mathrm{max}}(1.5) \approx  .5365$ 
and  $\gamma_{\mathrm{max}}(2) \approx  .2765$ in a very good agreement with
Fig.~\ref{spectrum_disk}a) and b). 
This bound corresponds to the imaginary part of zeros of $s_m(x)$ with $m=0$
and large $x$ and can be proved from semiclassical approximation to
(\ref{Sm}).  Physically, it is natural that all the states have a leakage
smaller than a particle moving along the diameter of the circle.  

As was mentioned above, the resonances in Figs.~\ref{spectrum_disk}a) and b)
are not the only existing quasi-bound states of the dielectric disk. It is
known (see e.g. \cite{nous}) that functions $s_m(x)$ have another series
of zeros corresponding to the shape (or outer) resonances,
$\tilde{x}_{m,p}$, also depending on  2 quantum numbers $m$ and $p$.  
\begin{figure}
\begin{center}
\includegraphics[width=.5\linewidth,angle=-90]{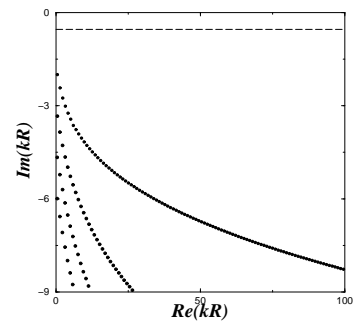}
\end{center}
\caption{Outer (shape) resonances for the dielectric circular cavity with
  $n=1.5$ Dashed line indicates the lowest bound (\ref{limit}) for
  imaginary parts of resonances in Fig.~\ref{spectrum_disk}a)}  
\label{outer}
\end{figure}
At Fig.~\ref{outer} some of these resonances are plotted. Notice the
difference in the vertical scale of this figure in comparison with
Figs.~\ref{spectrum_disk}a) and b). All resonances at
Fig.~\ref{spectrum_disk}a) are situated above the dashed line in
Fig.~\ref{outer}.  In \cite{nous} it was shown that  positions of these
outer resonances are described asymptotically by the equation 
\begin{equation}
\dfrac{H_m^{(1)\prime}}{H_m^{(1)}}(\tilde{x}_{m,p})
=\mathrm{i}\sqrt{n^2-\frac{m^2}{\tilde{x}_{m,p}^2}}  \ . 
\end{equation}
Its solutions with the smallest modulus of the imaginary part have
$\tilde{x}_{m,p}\approx m$ and are close to complex zeros of the Hankel
function in the denominator of this equation. In \cite{nous} it is shown
that in this case   
\begin{equation}
\tilde{x}_{m,p}=z_{m,p}-\frac{\mathrm{i}}{\sqrt{n^2-m^2/z_{m,p}^2}}
+\mathcal{O}(\mathrm{Im}\, z_{m,p}/\mathrm{Re}\, z_{m,p}) 
\label{outer_zeros}
\end{equation} 
where $z_{m,p}$ is a complex  zero of $H_m^{(1)}(z)$ function. It is known
\cite{bateman} that  at the principal Riemann sheet  of $H_m^{(1)}(z)$ cut
along the negative part of the real axis  such zeros have negative imaginary
parts,  lie symmetrically  with respect to the imaginary axis, and for
integer $m$ there is $[m/2]$ zeros with positive real part. These Hankel
zeros with the lowest imaginary part can asymptotically be computed by the
expansion (see e.g. \cite{nous})  
\begin{equation}
z_{m,p}= m+\eta_p\left ( \frac{m}{2}\right )^{1/3}\mathrm{e}^{-2\pi
  \mathrm{i}/3}+\mathcal{O}(\eta_p^2/m^{1/3}) 
\end{equation}
and $\eta_p$ is the modulus of the $p^{\mathrm{th}}$ zero of the Airy
function ($p=1,2,\ldots$). With a good precision  these Airy zeros are
described by the semiclassical formula 
\begin{equation}
\frac{2}{3}\eta_p^{3/2}\approx \pi(p-\frac{1}{4})\ 
\end{equation} 
which works well even for small $p$. In general, shape resonances of
dielectric circular cavity  are situated at $\mathcal{O}(1)$ distances from
shape resonances of circular cavity with Dirichlet boundary conditions.  

From the above arguments  it follows that the additional resonances have
quite large imaginary parts and are located below the curve 
\begin{equation}
-\mathrm{Im}\,x<\mathrm{const}\  |\mathrm{ Re }\,x|^{1/3}
\label{bound}
\end{equation}  
so they are well separated from the main resonances obeying
(\ref{max_gamma}). This bound has been  proved \cite{trapping_orbits},
\cite{zworski_sjostrand} for the scattering on any smooth $(C^{\infty})$
convex cavity provided no  
trapping orbits exist. If the cavity boundary is not smooth only the  weaker
bound is known \cite{martinez} 
\begin{equation}
-\mathrm{Im}\,x <\mathrm{const}\  \log |\mathrm{ Re }\,x|\ .
\end{equation} 
 
 \section{Weyl's law}\label{weyl_term} 
 
The circular cavity is useful to find the Weyl law (\ref{new_weyl}) for the
smooth part of the counting function. As usual (see e.g. \cite{uzy}), it can
be extracted by considering    pure imaginary  values  of momentum
$k=\mathrm{i}s$ and calculating the asymptotics  of the Bessel functions
from the known formulas (see e.g. \cite{bateman}). In this manner  one gets 
\begin{eqnarray}
&&\mathrm{i}
s_m(\mathrm{i}s)=\sqrt{\frac{m^2}{s^2}+n^2}+\sqrt{\frac{m^2}{s^2}+1}\\ 
&-&\frac{n^2 s}{2(m^2+n^2 s^2)}+\frac{s}{2(m^2+s^2)}  +\mathcal{O}(s^{-2})\ .
\nonumber
\end{eqnarray} 
Changing the summation over $m$ in (\ref{d_circle}) into the integration
from $-\infty$ to $\infty$  one obtains that  two main terms of the smooth
part of the right hand side of (\ref{d_circle}) have the following form 
\begin{equation}
\langle \frac{\partial\ }{\partial E} \ln \det \ \mathbf{S}(E)  \rangle= 
\frac{A}{4\pi }(n^2-1) +[r(n)-1]\frac{L}{8\pi k}+\mathcal{O}(k^{-3})\ 
\label{asymp_1}
\end{equation}
where $r(n)$ is defined in (\ref{new_weyl}).

But this formula gives the asymptotic behavior of  the determinant of the
full $\mathbf{S}$-matrix which is not necessarily   related directly  with
cavity resonances.   According to (\ref{matS}) the $\mathbf{S}$-matrix for
the dielectric circular cavity is the product of 2 factors 
\begin{equation}
\mathbf{S}_m(E)=\mathbf{S}_m^{(0)}(E)\mathbf{S}_m^{(1)}(E)
\label{product}
\end{equation} 
where 
\begin{equation}
\mathbf{S}_m^{(0)}(E)=-\frac{H^{(2)}_m(x)}{H^{(1)}_m(x)}
\end{equation}
and 
\begin{equation}
\mathbf{S}_m^{(1)}(E)=\frac{s_m^*(x)}{s_m(x)}\ .
\end{equation}
It is easy  to check that the $\mathbf{S}^{(0)}$-matrix is the
$\mathbf{S}$-matrix  for the scattering on a circular disk with the
Dirichlet boundary conditions whose asymptotic behavior is known (see
e.g. \cite{uzy} and references therein) 
\begin{equation}
\langle \frac{\partial\ }{\partial E} \ln \det \ \mathbf{S}^{(0)}(E)  \rangle=
-\frac{A}{4\pi } -\frac{L}{8\pi k}+\mathcal{O}(k^{-3})\ .
\label{asymp_2}
\end{equation}
The matrix element $\mathbf{S}_m^{(1)}(E)$ is the ratio of 2 functions:
$s_m(x)$ defined in (\ref{Sm}) and its complex conjugate. It is convenient
to rewrite it as the ratio of two other functions 
\begin{equation}
\mathbf{S}_m^{(1)}(E)=\frac{F_m^*(x)}{F_m(x)}\ 
\end{equation}
where
\begin{equation}
F_m(x)=\left [
  nJ_m^{\prime}(nx)-J_m(nx)\frac{H^{(1)\,\prime}_m(x)}{H^{(1)}_m(x)}\right ] 
x^{m-1}\ .
\end{equation}
The factor $x^{m-1}$ is introduced to insure that $F_m(x)$ tends to a
constant independent on $x$ when $m\to \infty$. 

 The function $F_m(x)$ has 2 groups of zeros (cf.  (\ref{zeros})). The first
 includes all usual resonances which obeys the inequality (\ref{max_gamma})
 and the second consists of additional resonances (shape resonances)  with
 large imaginary part (\ref{bound}). But it has also a series of poles
 coming from zeros of the Hankel function $H_m^{(1)}(x)$.   According to
 (\ref{outer_zeros})   the position of shape resonances in two leading
 orders differs from the corresponding zero of the Hankel function only by a
 finite shift.  From the bound (\ref{bound}) it follows also that for a
 finite $\gamma>0$ there is only a finite number of additional resonances
 whose imaginary part  obeys the inequality $\mathrm{Im}k<-\gamma$. 

 These two arguments demonstrate that with a big precision additional
 resonances and poles of the Hankel functions corresponding to shape
 resonances of the external Dirichlet problem cancel each other.    It means
 that function $F_m(x)$ can be considered as having effectively only
 resonances with small imaginary parts. Consequently,   the mean value of  
$(\ln \det \ \mathbf{S}^{(1)})^{\prime}$ gives the average value of the
counting function of resonances. From (\ref{product}) it follows that the
later  equals to the difference between (\ref{asymp_1}) and (\ref{asymp_2})  
\begin{eqnarray}
&&\langle \frac{\partial\ }{\partial E} \ln \det \ \mathbf{S}^{(1)}(E)  \rangle\\
&=&\langle \frac{\partial\ }{\partial E} \ln \det \ \mathbf{S}(E)  \rangle -
\langle \frac{\partial\ }{\partial E} \ln \det \ \mathbf{S}^{(0)}(E)  \rangle
\nonumber
\end{eqnarray}
which finally leads to  the asymptotics (\ref{weyl_dielc}). A careful
separation of the internal and external resonances in higher order corrections
remains an open problem.  

The fact that in order to find the smooth part of the resonance counting
function for 
a dielectric cavity in the leading  order one has  to divide the determinant
of the $\mathbf{S}$-matrix for the scattering on this  cavity  by  the
determinant of the $\mathbf{S}$-matrix for the scattering on the same cavity
but with the Dirichlet boundary condition may  physically be argumented as
follows. In the semiclassical limit a small part of a smooth boundary can be
considered as a straight segment. Resonances inside the cavity can be
determined from the knowledge of the reflection coefficient from this part
of the boundary as it is done in the Appendix~\ref{krein1dim}. It is easy to
see that the $\mathbf{S}$-matrix corresponding to the reflection from
outside the cavity differs just by a sign from the reflection coefficient
inside the cavity. But the  $\mathbf{S}$-matrix equals $-1$ is exactly the
$\mathbf{S}$-matrix for the scattering on the cavity with the Dirichlet
boundary conditions which explains the above statement.  

For a symmetric cavity it is  convenient to split resonances according to
their symmetry representations. Often it equivalent  to  consider a smaller
cavity  where certain boundaries are true dielectric boundaries but on
others parts of the boundary one may have in the simplest cases either
Dirichlet or Neumann boundary conditions. Now the total boundary
contribution to the mean counting function can be written as follows 
\begin{equation}
\frac{1}{4\pi}\left ( n(L_N-L_D)+r(n)L_0 \right )k\ .
\end{equation} 
Here $L_N$ and $L_D$ are the lengths of the boundary parts with respectively
Neumann and Dirichlet boundary conditions and $L_0$ is the length of the
true dielectric interface. 

\begin{figure}
\begin{minipage}{.49\linewidth}
\includegraphics[width=.95\linewidth]{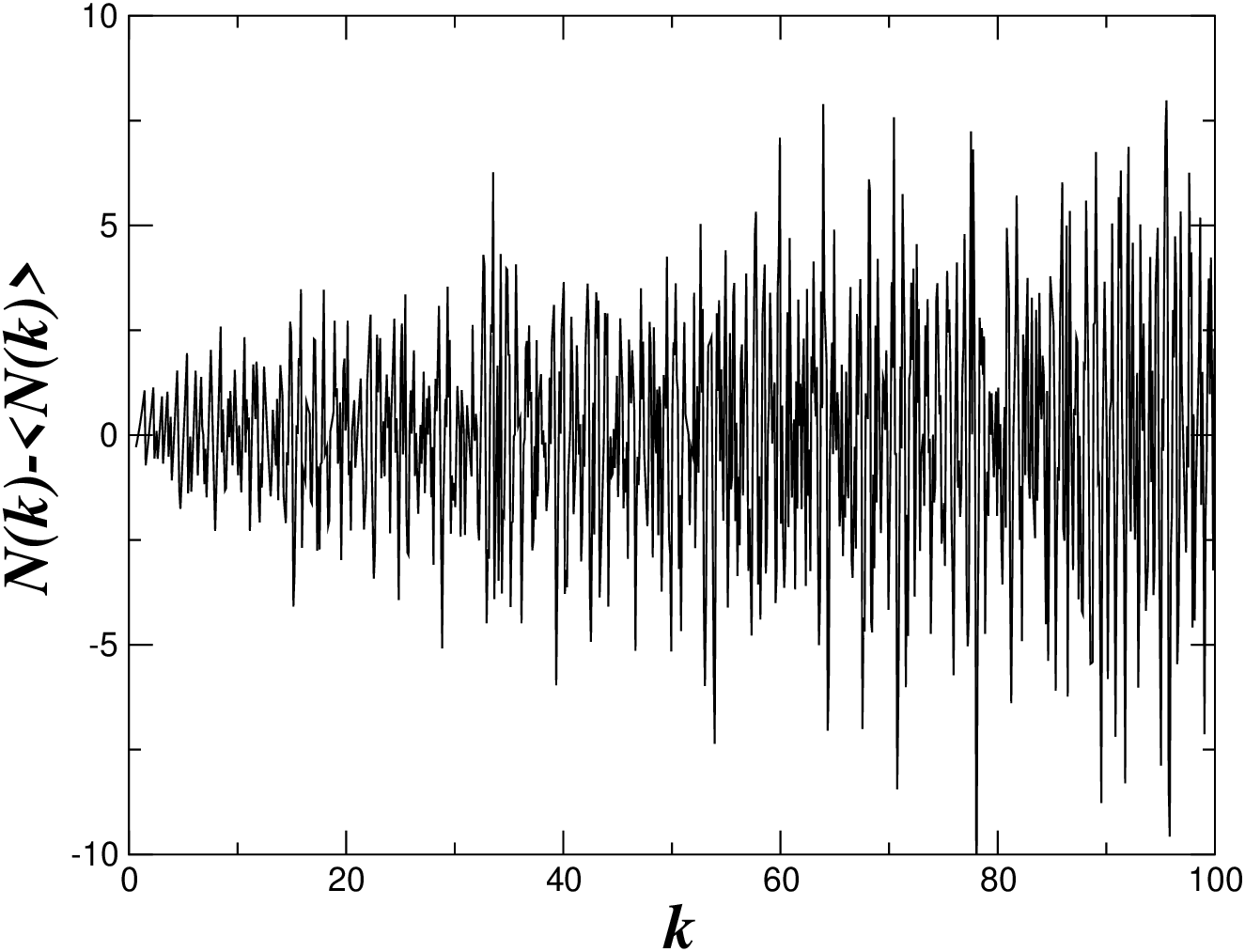} 
\begin{center}a)\end{center}
\end{minipage}
\begin{minipage}{.49\linewidth}
\includegraphics[width=.95\linewidth]{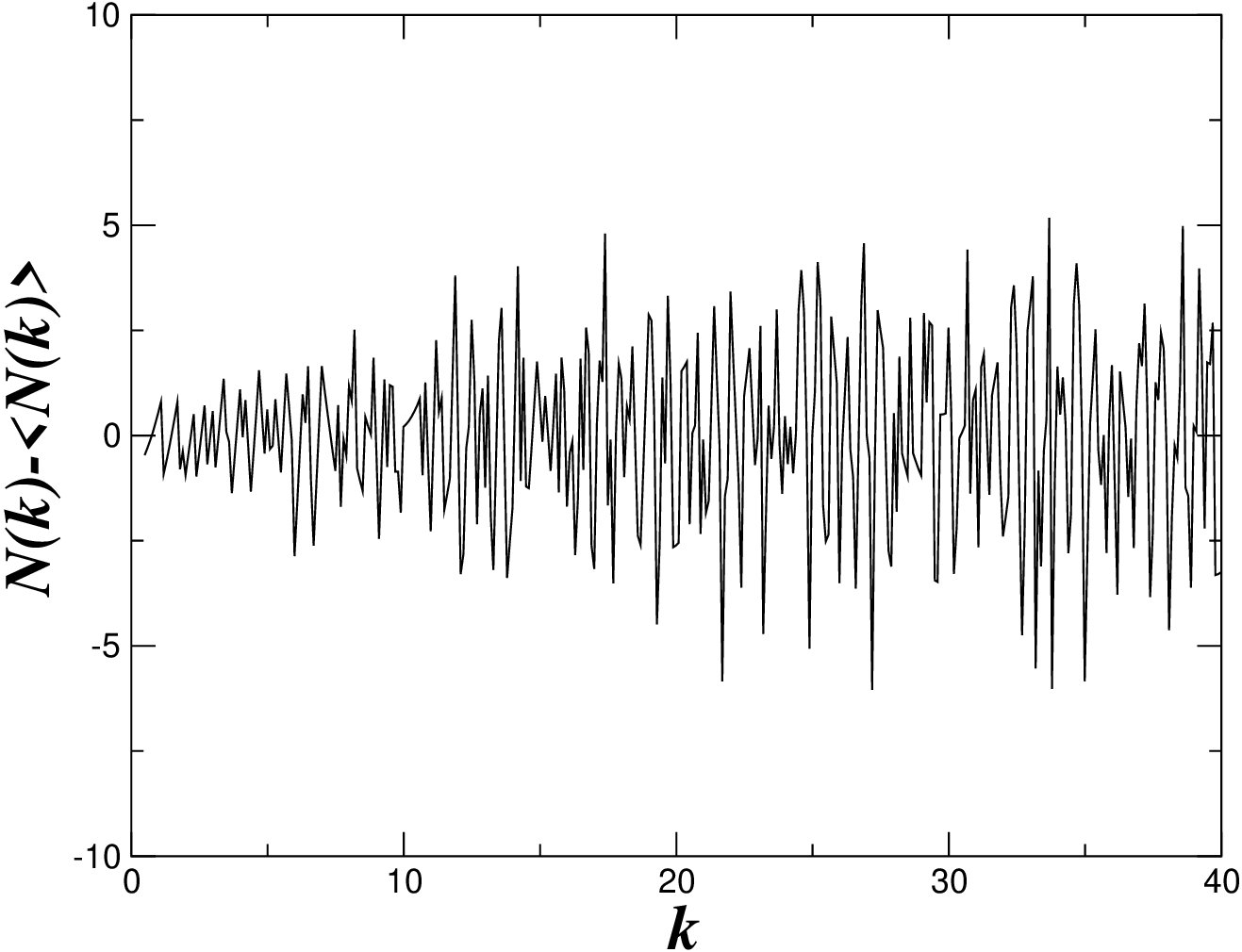} 
\begin{center}b)\end{center}
\end{minipage}
\caption{Difference between the exact counting functions of 
resonances of circular cavity with unit radius and  their  quadratic  fits (\ref{n_1.5}) and (\ref{n_2}): (a)   for $n=1.5$, (b) for $n=2$. }
\label{weyl_law}
\end{figure}

 The knowledge of the spectrum for dielectric disk
 (cf. Fig.~\ref{spectrum_disk}) permits us to check the derived formula for
 the mean counting function. First we count how many resonances  exist with
 real part less than a fixed value,  Re $k_m<k$. These resonances have to be
 taken into account with their multiplicities. For the circular cavity
 states with $m\neq 0$ are double degenerated and states with $m=0$ are
 simple.  Then  the resulting curves are fit by a polynomial of second
 degree with the highest term equals $n^2Ak^2/4\pi$. As for  $R=1$, $A=\pi$
 and $L=2\pi$, one gets the following fits: 
 \begin{itemize}  
\item For $n=1.5$:  
\begin{equation}
\bar{N}(k)=\frac{2.25}{4}k^2+\mathbf{.5139}k+.1518\ .
\label{n_1.5}
\end{equation} 
\item For $n=2$: 
\begin{equation}
\bar{N}(k)= k^2+\mathbf{.5446}k+.034\ .
\label{n_2}
\end{equation}
\end{itemize}
The quality of the fits can be seen from Fig.~\ref{weyl_law} where the differences between numerically calculated counting functions and the above fits are plotted.  
The theoretical prediction for the coefficient of the linear in $k$ term  is
$\tilde{r}(n)/2$ where $\tilde{r}(n)$ is defined in
(\ref{new_weyl}). Numerically one finds that these coefficients should be 
equal to $\mathbf{.5125}$ for $n=1.5$ and to $\mathbf{.5420}$ for $n=2$ in a
good agreement with the above numerical fits.  
   
\section{Periodic orbit contribution}\label{periodic}

To get oscillating contributions to the trace formula for a circular cavity
we start from the expression (\ref{d_circle}). Expressing the Bessel
function in (\ref{Sm}) through the Hankel functions of the first and the
second kinds $J_m(x)=(H_m^{(1)}(x)+H_m^{(2)}(x))/2$ 
and performing straightforward transformations of $s_m(x)$ in (\ref{Sm}) one
gets a formal series for the density of states 
\begin{eqnarray}
&&d(E)-d_0(E)=\frac{R^2}{2\pi x}\sum_{m=-\infty}^{\infty}\mathrm{Im}\,\left
  [ \frac{n^2-1}{B_m(x)}\right .\\ 
&&+\left .\frac{4\mathrm{i}}{\pi
    x}\sum_{r=1}^{\infty}P_m(x)[R_m^{(\mathrm{e})}(x)]^rE_m^r(x)\right ] . 
\nonumber
\end{eqnarray}
Using the Poisson summation formula, the sum over $m$ can be substituted by
the integration over $m$ and the summation over a conjugate integer $M$ 
\begin{eqnarray}
&&d(E)-d_0(E)=\frac{R^2}{4\pi
  \mathrm{i}x}\sum_{M=-\infty}^{\infty}\int_{-\infty}^{\infty}
\mathrm{d}m\,\mathrm{e}^{2\pi \mathrm{i}M m} \label{integral}\\    
&&\times \left [ \frac{n^2-1}{B_m(x)}+
\frac{4\mathrm{i}}{\pi x}\sum_{r=1}^{\infty}P_m(x) [R_m^{(\mathrm{e})}(x)]^r
E_m^r(x)\right ]+ \mathrm{c.c.} 
\nonumber
\end{eqnarray}
Here the following notations have been introduced
\begin{equation}
E_m(x)=\frac{H_m^{(1)}(x)}{H_m^{(2)}(x)}\ ,
\end{equation}
\begin{equation}
R_m^{(\mathrm{e})}(x)=-\frac{A_m(x)}{B_m(x)}\ ,
\label{exact}
\end{equation}
\begin{equation}
A_m(x)=n\frac{H_m^{(1)\prime }}{H_m^{(1)}}(nx)-\frac{H_m^{(1)\prime
  }}{H_m^{(1)}}(x)\ , 
\end{equation}
\begin{equation}
B_m(x)=n\frac{H_m^{(2)\prime }}{H_m^{(2)}}(nx)-\frac{H_m^{(1)\prime }}{H_m^{(1)}}(x)\ ,
\end{equation}
and 
\begin{equation}
P_m(x)=(n^2-1)\left (H_m^{(1)}(nx) H_m^{(2)}(nx) A_m(x) B_m(x)\right )^{-1}.
\label{prefactor}
\end{equation}
No approximation has been done in Eq.~(\ref{integral}) and it can serve as a
starting point for  derivations of dominant terms and corrections to
them. In the semiclassical limit $x\to\infty$ it can be simplified  by using
standard asymptotics of the Hankel function \cite{bateman} 
\begin{eqnarray}
&&H_m^{(1)}(x)\simeq
\frac{\sqrt{2/\pi}}{(x^2-m^2)^{1/4}}\mathrm{e}^{\mathrm{i}\phi_m(x)}\nonumber\\ 
&&\times\left [1-\frac{b_1(x)}{\sqrt{x^2-m^2}}+\mathcal{O}(x^{-2})\right ] 
\label{hankel}
\end{eqnarray}
where 
\begin{equation}
 \phi_m(x)=\sqrt{n^2 x^2 -m^2}-m \arccos \left( \frac{m}{nx} \right) -
 \frac{\pi}{4} 
 \end{equation}
and
\begin{equation}
b_1(x)=\frac{1}{8}-\frac{5}{24}(1-x^2/m^2)^{-1}\ .
\end{equation} 
In this manner one concludes that, when  $x\to \infty$ and $|m|<nx$,  
$E_m(x)\to \mathrm{e}^{2\mathrm{i}\phi_m(x)}$, $P_m(x)\to
\sqrt{n^2-m^2/x^2}$, and  
\begin{equation}
R_m^{(\mathrm{e})}(x)\to R_m(x)= \frac{\sqrt{n^2-m^2/x^2}-\sqrt{1-m^2/x^2}} 
{\sqrt{n^2-m^2/x^2}+\sqrt{1-m^2/x^2}}\ .
\end{equation}
From these equations it follows that the dominant periodic orbit contribution is 
\begin{eqnarray}
&&d^{(osc)}(E)=\frac{R^2}{4\pi
  x}\sum_{M=-\infty}^{\infty}\sum_{r=1}^{\infty}\label{poisson}\\ 
&&\times \int_{-\infty}^{\infty}\mathrm{d}m \sqrt{n^2-\frac{m^2}{x^2}}R_m^r(x)
\mathrm{e}^{\mathrm{i} S_{M,r}(m)}+\mathrm{c.c.}
\nonumber
\end{eqnarray} 
where the action (the phase of the exponent) 
\begin{equation}
S_{M,r}(m)=2\pi  Mm+2r\phi_m(x)\ .
\label{action}
\end{equation}
The last step, as usual, consists in the computation of the integral over
$m$ by the  
stationary phase method. The saddle point value of $m=m_{sp}$ is determined
from the equation 
\begin{equation}
\frac{\partial }{\partial m}S_{M,r}(m)=0
\label{saddle_equation}
\end{equation}
whose solution 
\begin{equation}
m_{sp}=nx\cos \theta_{M,r}
\label{saddle_point}
\end{equation}
with
\begin{equation}
\theta_{M,r}=\pi \frac{M}{r}
\end{equation}
corresponds geometrically to the periodic orbit of  the circular cavity in
the shape of regular polygon with $r$ vertices going around the center $M<r$
times. If $M$ and $r$ are co-prime integers the periodic orbit is
primitive. Otherwise, it corresponds to a primitive periodic orbit repeated
$d$ times where $d$ is the largest common factor of $M$ and $r$.  

The action (\ref{action}) can be expanded into a series of deviation from
from the saddle point value $m=m_{sp}+\delta m$.  One gets  
\begin{equation}
S_{M,r}(m_{sp}+\delta m)=n kl_p -\frac{\pi}{2}r+\frac{g_2}{2}\delta m^2
+\mathcal{O}(\delta m^3) 
\label{higher_terms}
\end{equation}
where  $l_p$ is the periodic orbit length  
\begin{equation}
l_p=2r R\sin\theta_{M,r}
\label{lp}
\end{equation}
and $g_2$ is   the second derivative of the action computed at the saddle point 
\begin{equation}
g_2=\frac{2r}{nx \sin \theta_{M,r}}\ .
\label{second_der}
\end{equation}
Computing the integral over $m$ by the saddle point approximation one gets
the expected result (\ref{dE_int})  
\begin{equation}
d^{(osc)}(E)=\frac{n^{3/2}}{\pi}
\sum_{M,r}
\frac{\mathcal{A}_p}{\sqrt{2\pi  k l_p}} R_p^r 
\mathrm{e}^{\mathrm{i}[nkl_p-r\pi/2+\pi/4]} +\mathrm{c.c.}
\label{expected}
\end{equation}
where 
\begin{equation}
\mathcal{A}_p=\pi R^2\sin^2 \theta_{M,r}
\label{Ap}
\end{equation}
is the area occupied by the considered periodic orbit family and $R_p$ is
the Fresnel reflection coefficient (\ref{reflection}) calculated at the
incidence angle  
\begin{equation}
\theta=\frac{\pi}{2}-\theta_{M,r}\ .
\label{incidence_angle}
\end{equation}
As classical motion inside a circle conserves the angle with the boundary,
the mean value in (\ref{dE_int}) equals the product of reflection
coefficients in all incident points. 

\begin{figure}
\begin{minipage}{.49\linewidth}
\begin{center}
\includegraphics[width=.95\linewidth]{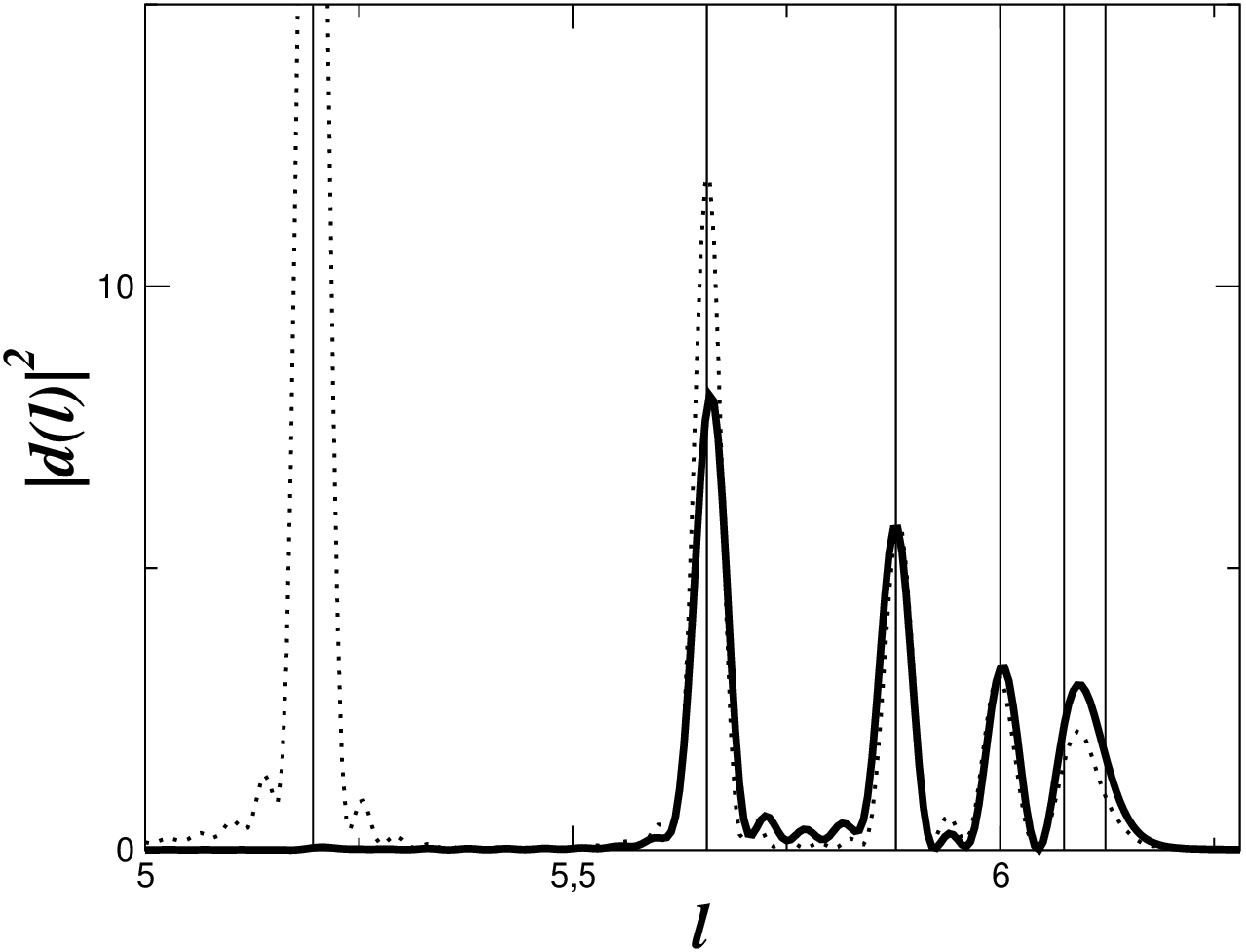} 
\end{center}
\begin{center}a)\end{center}
\end{minipage}\hfill
\begin{minipage}{.49\linewidth}
\includegraphics[width=.95\linewidth]{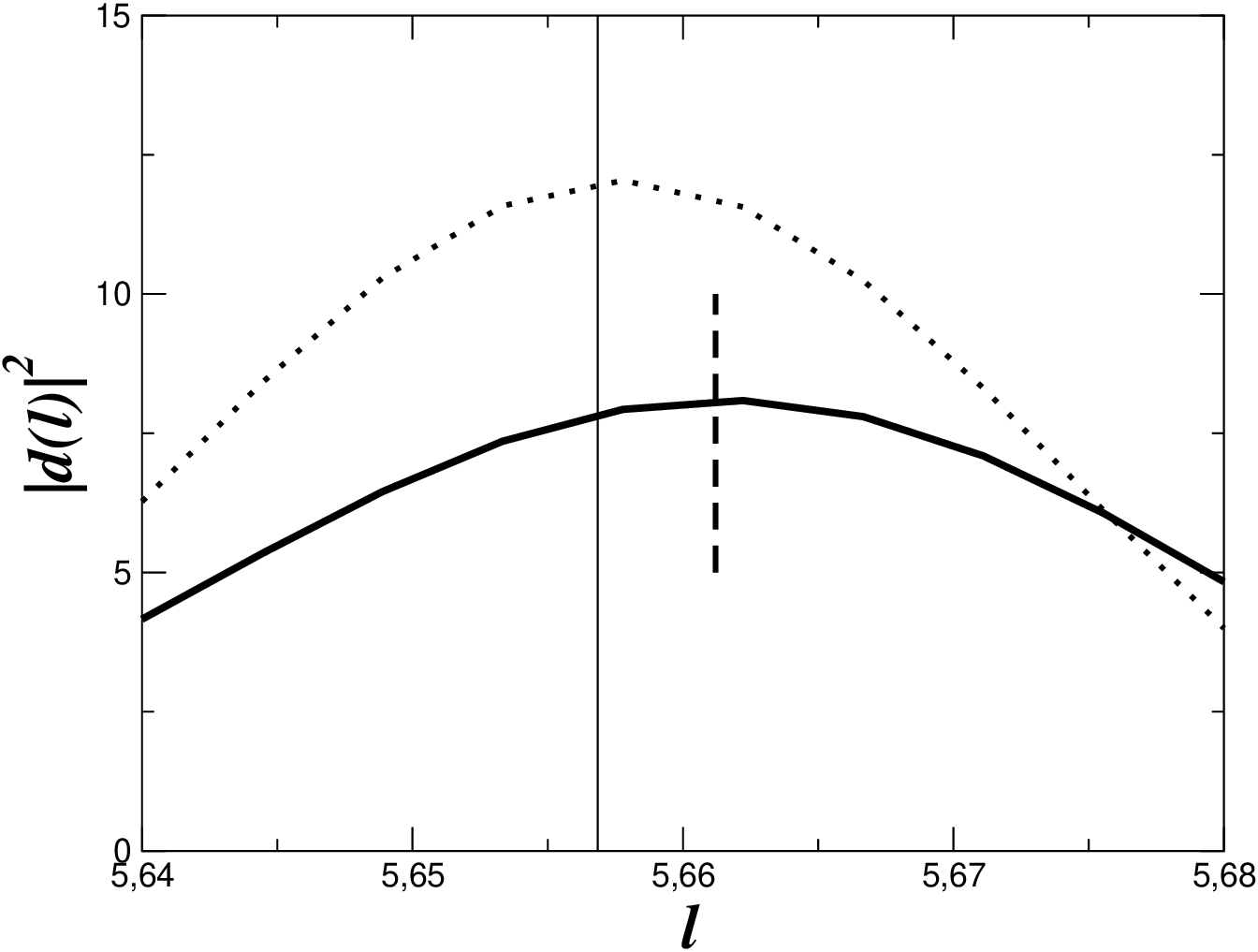} 
\begin{center}b)\end{center}
\end{minipage}
\caption{(a) Solid line:  length density for dielectric circular cavity with
  $n=1.5$ and $R=1$.  Dotted  
 line: the same but for  circular cavity with the Dirichlet boundary conditions.
 (b) Goos-Hänchen shift associated with the square periodic orbit. Solid
 line corresponds to the dielectric disk and the dotted line represents the
 circular closed cavity. Small vertical line indicates the maximum of length
 spectrum for dielectric disk. } 
\label{cercle}
\end{figure}

At Fig.~\ref{cercle}a) the length density (i.e. the Fourier transform of the
trace formula (\ref{fourier})) is presented for  the circular dielectric
cavity with the refraction coefficient equal to $1.5$. The vertical solid
lines indicate the lengths of  simplest periodic orbits. From the left to
the right these lines correspond to, respectively, triangle, square,
pentagon, hexagon,  heptagon, and octagon.  
For clarity the dotted line at this figure represents the length density for
the closed circular cavity with the Dirichlet boundary conditions.  As
expected the both models have peaks at periodic orbit lengths but their
amplitudes are different. In particular, as for $n=1.5$ the critical angle
is of the order of $42^{o}$,  triangular periodic orbit is not confined and
the corresponding peak for dielectric cavity is practically invisible.
Others peaks have similar heights. 
 
\section{Corrections to asymptotic regime}\label{goos}

According to (\ref{dE_int}) the contribution of a periodic orbit for an open
dielectric cavity differs from a closed cavity only by the value of the
total reflection coefficient. Thus, when all incidence angles for a periodic
orbit are greater than the critical angle (\ref{critical}), the modulus of
its contribution to the trace formula has to be the same as for the closed
cavity. The curves at Fig.~\ref{cercle} have been normalized in such a way
that  their peaks at pentagonal periodic orbit  are the same heights.  But
then the amplitudes corresponding to the square orbit are clearly different
(dielectric cavity peak is of the order of $0.7$ of the close cavity peak)
which contradicts the above prediction.  

Other important corrections are the so-called Goos-Hänchen shift \cite{g_h},
\cite{lai,martina}  which manifests itself as a slight difference of a
visible length from its geometric length for  an orbit reflected from a
dielectric interface and the Fresnel filtering \cite{fresnel_filt}, which
manifests itself as a slight shift of the angle of emission near the
critical angle for a dieletric boundary.  At Fig.~\ref{cercle}b) a region of
Fig.~\ref{cercle}a) close to the maximum of the length density around the
square periodic orbit is magnified. The peak of the close cavity represented
by dotted line is, as expected, centered at the geometric length of this
orbit ($l=4\sqrt{2}R\simeq 5.6568$ for $R=1$). But the peak associated with
the dielectric circular cavity is clearly shifted to the right. Numerically,
this shift is  small and for the considered interval of momentum it is of
the order of $0.004R$. 

The reason of these discrepancies is related with the fact that asymptotics
of the Hankel function (\ref{hankel}) is not valid when $m$ is close to $x$
(see e.g. \cite{bateman}).  The uniform approximation for large $m$ and all
$x$ is more complicated and is given e.g.  by the Langer formula
\cite{bateman} 
\begin{equation}
H_m^{(1)}(x)=\left (1-\frac{\arctan\, \omega}{\omega}\right  )^{1/2}
H_{1/3}^{(1)}(z)\mathrm{e}^{\mathrm{i}\pi/6}+\mathcal{O}(m^{-4/3})
\label{langer}
\end{equation}  
where $z=m(\omega-\arctan\, \omega )$ and $\omega=\sqrt{x^2/m^2-1}$.

Non-uniform character of this expression is clearly seen from  the following
asymptotics useful for further estimations  
\begin{eqnarray}
\frac{H_m^{(1)\prime}(x)}{H_m^{(1)}(x)}|_{x=m}&=&
\frac{\alpha}{m^{1/3}}+\mathcal{O}\left (\frac{1}{m}\right )\ ,\\ 
\frac{\partial }{\partial m}\frac{H_m^{(1)\prime}(x)}{H_m^{(1)}(x)}|_{x=m}&=&
\frac{\alpha^2}{m^{2/3}}+\mathcal{O}\left (\frac{1}{m^{4/3}}\right )\\
\frac{\partial^2 }{\partial m^2}\frac{H_m^{(1)\prime}(x)}{H_m^{(1)}(x)}|_{x=m}&=&\frac{2\alpha^3-2}{m}+\mathcal{O}\left (\frac{1}{m^{5/3}}\right)
\label{non_uniform}
\end{eqnarray}
where $\alpha=2^{-2/3}3^{5/6}\pi^{-1}\Gamma^2(2/3)\mathrm{e}^{2\pi \mathrm{i}/3}\simeq -.4592+.7954\mathrm{i}$.

To calculate carefully higher order corrections to the trace formula (\ref{expected}) is not an easy task. The main difficulty lies in the fact that the stationary phase method on which trace formulas are based on cannot, strictly speaking,   be applied when a periodic orbit has incident angles close to the critical one. The point is that in the stationary phase method it is implicitly assumed that non-exponential terms  are changed much slowly  than the action in the exponent. This is the case for the usual billiard systems where the whole perturbation series for the trace formula has been constructed \cite{gaspard_alonso}, \cite{vattay}. 

But as follows from (\ref{langer}) for dielectric cavities exact reflection
coefficient (\ref{exact}) near the critical angle changes at a scale of
$\delta \bar{m}\sim x^{1/3}$ 
but the action changes at a different scale of the order of $\delta
\tilde{m} \sim x^{1/2}$.  

When $x\to \infty$ the former is much smaller than the latter and the
stationary phase method when a trajectory hits a boundary with angle close
to the critical one can not be justified. To avoid this difficulty one can
use the Langer approximation (\ref{langer}) in the transitional region and
tabulate  the integrals without other approximations.  We will not perform
these computations here. Instead, we compare the saddle point result with
numerically calculated integral for the square and pentagonal periodic
orbits. For refractive index $n=1.5$ incident angles of the square periodic
orbit are close to  the critical angle but for the pentagonal orbit they are
far apart and the difference of these two cases will give the accuracy of
the pure saddle point results. 

From (\ref{integral}) it follows that the full contribution of a periodic orbit with fixed numbers $M$ and $r$ is given by the integral
\begin{equation}
d_p(E)=\mathrm{e}^{\mathrm{i} [nkl_p-r\pi/2+\pi/4]}J_{M,r}(x)
\end{equation}  
where 
\begin{eqnarray}
&&J_{M,r}(x)=\frac{R^2}{\pi^2x^2}\int_{-\infty}^{\infty} P_m(x)[R_m^{(\mathrm{e})}(x)]^rE_m^r(x)\nonumber\\
&&\times \mathrm{e}^{-\mathrm{i}[nkl_p-r\pi/2+\pi/4]}\mathrm{d}m
\label{exact_prefactor}
\end{eqnarray}
In the pure stationary phase approximation  integral (\ref{exact_prefactor}) equals
\begin{equation}
J_{M,r}^{(sp)}(x)=\frac{R^2}{\pi^2x^2} P_{m_{sp}}(x)[R_{m_{sp}}^{(\mathrm{e})}(x)]^r
\sqrt{\frac{2\pi}{g_2}}
\label{saddle_prefactor}
\end{equation}
which in the semiclassical limit tends to 
\begin{equation}
J_{M,r}^{(scl)}(x)=\frac{n^{3/2}\mathcal{A}_p}{\pi\sqrt{2\pi kl_p}}R_p^r
\label{semi_prefactor}
\end{equation}
as in (\ref{expected}). 

At the first sight it seems that a better approximation  consists in the
calculation of the prefactor (\ref{prefactor}) and the reflection
coefficient (\ref{exact}) in (\ref{saddle_prefactor}) directly at the saddle
point (\ref{saddle_point}) without the asymptotic formula
(\ref{hankel}). Roughly speaking, this is equivalent to use for the
reflection on a curved surface instead of the Fresnel diffraction
coefficient (\ref{reflection}) valid for an infinite plane interface the
reflection coefficient (\ref{exact}) computed at the saddle point value of
the incidence angle  $\theta$ as in (\ref{incidence_angle}). When this angle
is close to the critical angle from (\ref{exact}) one gets 
\begin{equation}
R_m(\theta,x)=\dfrac{n\cos \theta+\mathrm{i}H_m^{(1)\prime}(x)/H_m^{(1)}(x)}
{n\cos \theta-\mathrm{i}H_m^{(1)\prime}(x)/H_m^{(1)}(x)}
\label{schomerus}
\end{equation} 
where the value of the angular momenta $m=m(\theta)$ is related with angle $\theta$ as it follows from the saddle point relation (\ref{saddle_point})
\begin{equation}
m(\theta)=nx\sin \theta\ .
\end{equation}
This type of modification first has  been proposed in \cite{martina}. 

Nevertheless, we found that  this approximation numerically is not better
than the simple semiclassical formula (\ref{semi_prefactor}). This is
because the effective reflection coefficient (\ref{schomerus}) changes too
quickly close to critical angle. Therefore,  when a region which gives a
dominant contribution to the integral (\ref{exact_prefactor}) is close but
not equal to the saddle point in the exponent,  expression
(\ref{saddle_prefactor}) differs considerably from the correct one. For this
reason we prefer to compare  numerical values of the integral
(\ref{exact_prefactor}) not with (\ref{saddle_prefactor}) but with simpler
expression (\ref{semi_prefactor}). 

\begin{figure}
\begin{minipage}{.49\linewidth}
\begin{center}
\includegraphics[width=.89\linewidth,angle=-90]{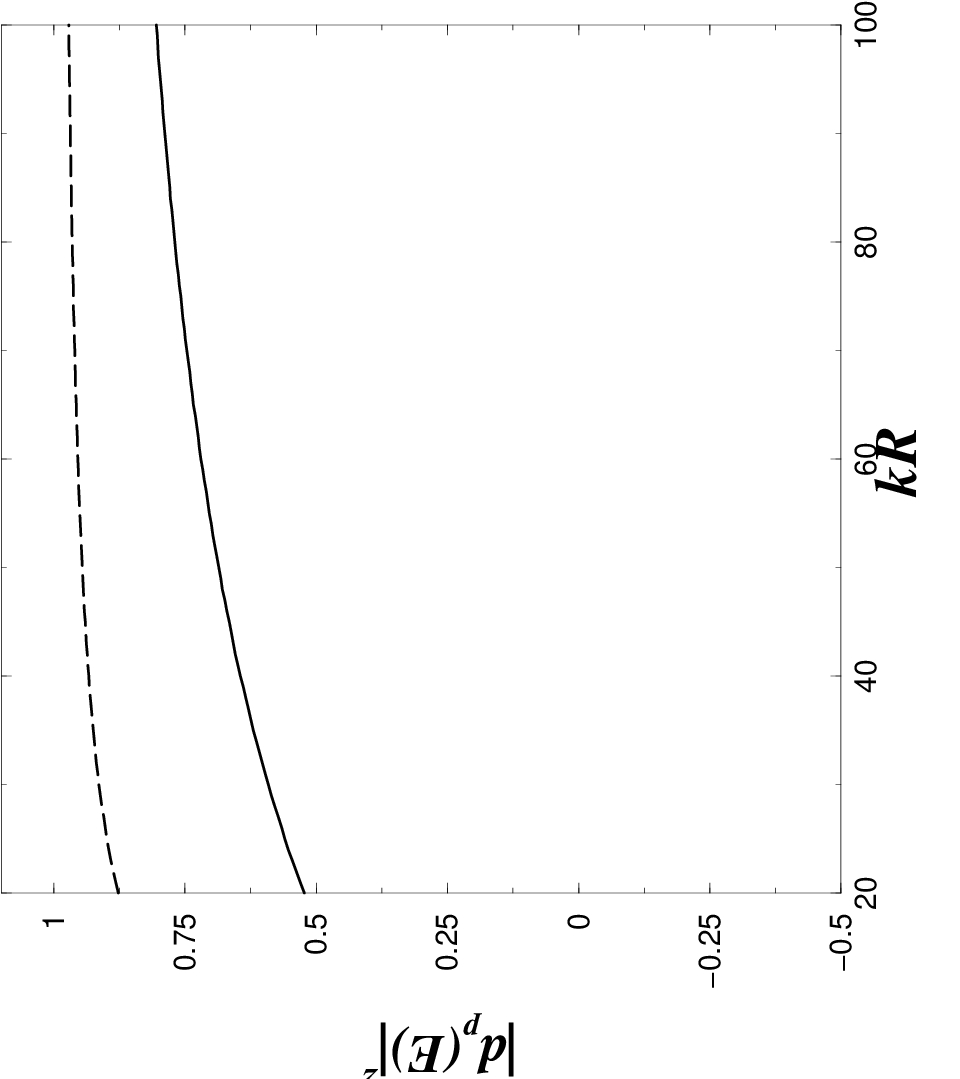} 
\end{center}
\begin{center}a)\end{center}
\end{minipage}\hfill
\begin{minipage}{.49\linewidth}
\begin{center}
\includegraphics[width=.88\linewidth,angle=-90]{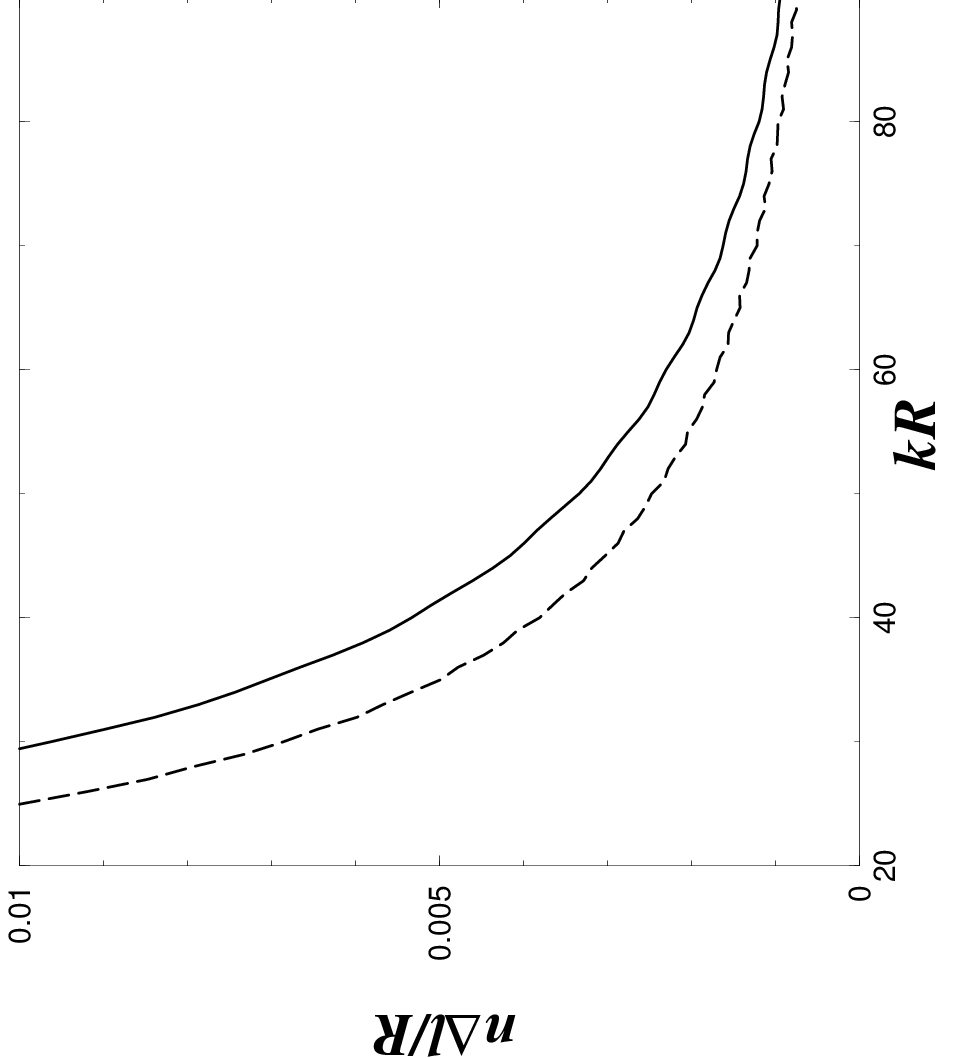} 
\end{center}
\begin{center}b)\end{center}
\end{minipage}
\caption{(a) Modulus square  of the integral (\ref{exact_prefactor}) divided by its semiclassical expression (\ref{semi_prefactor}). Solid line indicates the contribution of the square periodic orbit. Dashed line is the contribution of pentagonal orbit. (b) The derivatives of the phase of the same integral: solid and dashed lines  correspond respectively  to the square and pentagonal orbits. }
\label{goos_courbes}
\end{figure}

At Fig.~\ref{goos_courbes} the results of numerical calculations of the
integral (\ref{exact_prefactor}) for square and pentagonal periodic orbits
are presented. At Fig.~\ref{goos_courbes}a) the square of the modulus of
this integral divided by the semiclassical expression (\ref{semi_prefactor})
is given. As expected, contribution of the pentagonal periodic orbit is
well described by semiclassical formula. But the contribution of the square
periodic orbit differs considerably from the asymptotic result which
explains  quantitatively the observed differences of the square orbit
heights for closed and dielectric cavities seen at   Fig.~\ref{cercle}.  

The computed corrections permit also to explain the analog of the Goos-Hänchen shift  observed at Fig.~\ref{cercle}b). Let us write the periodic orbit contribution  in the form
\begin{equation}
d_p(E)=|d_p(E)|\mathrm{e}^{\mathrm{i}[knl_p-r\pi/2+\pi/4+\Psi_p(x)]}
\end{equation}
where the phase $\Psi_p(x)$ comes from the integral (\ref{exact_prefactor}). 
 
To compute the length spectrum  one has to calculate the Fourier harmonics of this expression (\ref{beta}). The shift of the peak  position, $\Delta l=l-l_p$ can be estimated from  the saddle point in the exponent
\begin{equation}
n\frac{\Delta l}{R}=\frac{\partial \Psi_p(x)}{\partial x}\ .
\label{formula}
\end{equation}
At Fig.~\ref{goos_courbes}b) the value of this derivative computed
numerically from the knowledge of (\ref{exact_prefactor}) is plotted for the
square and pentagonal orbits. For the square orbit its value is larger than
for pentagonal one and is of the same order as the  observed shift in
Fig.~\ref{cercle}b).   As we are looking for a very small effect, mutual
influence of different periodic orbits has to be taken into account when a
precise determination of the Goos-H\"anchen shift is required.

\section{Conclusion}

The goal of the paper is the construction of the trace formula for open
dielectric cavities. By using the Krein formula it was demonstrated that
such trace formula can be written in the form (\ref{dk}) 
\begin{eqnarray}
&-&\frac{1}{\pi}\sum_{m}\dfrac{\im E_m}{(k-\mathrm{Re}\, E_m)^2+(\mathrm{Im}\, E_m)^2}\nonumber\\
&=&\bar{d}(E)+\sum_{p}\big (d_p(E)+d_p^*(E)\big )
\label{trace_trace}
\end{eqnarray}
where in the left-hand side the sum is taken over all internal resonances
and in the right-hand side the summation includes all periodic orbits for
the free motion inside the cavity. The term $\bar{d}(E)$ is the mean density
of these resonances given by the derivative of  (\ref{weyl_dielc}). Its two
main terms are  
 \begin{equation}
 \bar{d}(E)=\frac{An^2}{4\pi}+\tilde{r}(n)\frac{L}{8\pi k}
\end{equation}
where $k=\sqrt{E}$, $A$ and $L$ are the area and the perimeter of the
cavity, and   $\tilde{r}(n)$ is as follows 
\begin{equation}
 \tilde{r}(n)=1+\frac{n^2}{\pi}\int_{-\infty}^{\infty}\frac{\mathrm{d}t}{t^2+n^2}R(t)-
 \frac{1}{\pi}\int_{-\infty}^{\infty}\frac{\mathrm{d}t}{t^2+1}R(t)
\label{tilde}
\end{equation}
where $R(t)$ is the reflection coefficient at imaginary momentum  defined in
(\ref{new_coeff}). 
Terms related with $R(t)$ correspond to the reflections from inside and
outside the boundary but $1$ in (\ref{tilde}) appears due to the necessity
of removing the $\mathbf{S}$-matrix for the scattering on the cavity with
the Dirichlet boundary conditions.  

The contribution from a particular periodic orbit, $d_p(E)$, to the trace
formula (\ref{trace_trace}) differs from the same closed billiard only by
the product over  Fresnel reflection coefficients (\ref{reflection})
computed at all reflection points as in (\ref{dE_chao}) for isolated orbits
or their average over a periodic orbit family as in  (\ref{dE_int}).  

We have  discussed also corrections to the above asymptotic semiclassical
formulas. Due to rapid changes of the reflection coefficient close to the
critical angle corrections for dielectric cavities are larger than for
closed cavities. In particular, higher order corrections permit to estimate
the analog of the Goos-Hänchen shift for the peak position in dielectric
cavity length density.  

Obtained formulas were compared with numerical calculation of resonances for
the circular dielectric cavity and in all cases a  good agreement was
observed.  

\acknowledgments

The authors are grateful to M. Lebental whose experimental results stimulate
the interest to the study of dielectric cavities. It is a pleasure to thank
M. Hentschel, S. Nonnenmacher,  U. Smilansky, M. Zworski and J. Wiersig for
numerous useful discussions.  

\appendix

\section{Krein formula for   one dimensional  systems}\label{krein1dim}

To illustrate the Krein formula in the simplest case of a one-dimensional system consider a particle moving along the positive half-axis in a potential without bound states. For simplicity, assume that (i) $\psi(0)=0$ and (ii) the potential is zero for distances larger than $a$. The corresponding quantum problem has only continuous spectrum. To get artificially the discrete spectrum let us impose that at some large distance   $R\gg a$ the wave function obeys 
\begin{equation}
\psi(R)=0\ .
\label{dirichlet}
\end{equation}
Eigenfunctions of a such  system at large distances $x>a$ have the form of plane waves 
\begin{equation}
  \psi(x)\propto \sin(k x+\delta(E)), 
\label{psi1d1}
\end{equation}
where $k=\sqrt{E}$ and the phase shift $\delta(E)$ contains all information about the potential.

Eigenmomenta, $k_m$,   of this system are determined from the condition  (\ref{dirichlet})  and they read 
\begin{equation}
  k_m R+\delta(E)=\pi m
\label{k1d}
\end{equation}
where integer $m$ counts the number of states. By definition, the density of states, $d(E)$,  is the number of states within an interval $\mathrm{d}E$
\begin{equation}
  d(E)\equiv \frac{\ud m}{\ud E}=\frac{R}{2\pi\sqrt{E}}+\frac{1}{\pi}\frac{\partial
    \delta(E)}{\partial  E} \ .   
\label{de1d}
\end{equation}
But the function (\ref{psi1d1}) with $\delta(E)=0$ is the eigenfunction of the free problem without the potential. Therefore, the first term in the last equation describes the free density of states
\begin{equation}
d_0(E)=\frac{R}{2\pi\sqrt{E}}\ .
\end{equation}    
The excess density due to the presence of the potential is the difference between the level density with the potential and without it. From (\ref{de1d}) it follows that 
\begin{equation}
d(E)-d_0(E)=\frac{1}{\pi}\frac{\partial \delta(E)}{\partial  E} \ .  
\end{equation}
Instead of the phase shift it is instructive to introduce    the
$\mathbf{S}$-matrix for the scattering on the potential defined as the ratio
of the amplitude of the outgoing wave to the amplitude of the incoming
wave. For the function (\ref{psi1d1}) one has    
\begin{equation}
  \mathbf{S}(E)= -\mathrm{e}^{2\ic \delta(E)}\ .
\label{s1d}
\end{equation}
Therefore,  the excess density of states reads
\begin{equation}
  d(E)-d_0(E)=\frac{1}{2\pi\ic} \frac{\partial \ln \mathrm{S}(E)}{\partial E}\ ,
\label{krein1d}
\end{equation}
which is the 1-dimensional version of the Krein formula (\ref{fkrein}). 

\begin{figure}
\begin{center}
\includegraphics[width=.7\linewidth]{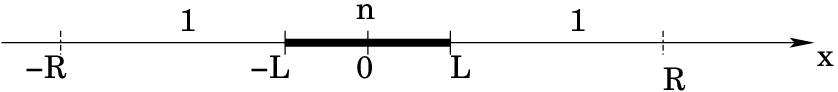} 
\caption{1-dimensional dielectric cavity. Refractive indices are indicated above the axis.}
\label{bardiel}
\end{center}
\end{figure}

It is also straightforward  to compute explicitly all necessary quantities for a 1-dimensional dielectric cavity \cite{these_Melanie}.   Consider a dielectric segment of size $L$ with the refractive index $n$ in a media with the refractive index $1$ as at  Fig.~\ref{bardiel}. 
After imposing condition (\ref{dirichlet}) one gets that the density of states $d(E)$
has to be determined from the formula
\begin{equation}
 \label{trg_d}
 d(E)=-\frac{1}{\pi}\int_0^R \im n^2(x)G(x,x)\ud x
\end{equation}
where $n(x)=n$ inside the cavity and $n(x)=1$ outside it, and  $G(x,y)$ is the Green function of our problem \cite{these_Melanie}. Using 
\begin{equation}
 d_0(E)=-\frac{1}{\pi}\int_0^R \im G^{(0)}_k(x,x)\ud
  x= \frac{R}{2\pi k}
\end{equation}
where $G^{(0)}(x,y)=\mathrm{e}^{\mathrm{i} k|x-y|}/2\mathrm{i}k$ is the free Green function after some calculations one finds that for antisymmetric states when $R\to \infty$
\begin{equation}
  \label{dens1d}
  d(E)-d_0(E)=-\frac{L}{2\pi k}+\frac{nL}{2\pi k}\frac{1-r^2}{|1+r \mathrm{e}^{2\ic
      n k L}|^2} \ 
\end{equation}
where $r$ is the Fresnel reflection coefficient from dielectric boundary
\begin{equation} 
r=\frac{n-1}{n+1}\ .
\end{equation}
Correspondingly,  the $\mathbf{S}$-matrix for antisymmetric states is
\begin{equation}
\label{matS1d}
\mathbf{S}(E)=-\mathrm{e}^{2\ic (n-1)kL} 
\frac{1+r \mathrm{e}^{-2\ic n k L}}{1+r \mathrm{e}^{2\ic n k L}}\ ,
\end{equation}
which, of course, agrees with 1-dimensional Krein formula  (\ref{krein1d}). This $\mathbf{S}$-matrix  has an infinite number of poles (or resonances) corresponding  to the solutions of the equation 
$1+r \mathrm{e}^{2\ic n k_m L}=0$ or explicitly  
\begin{equation}
k_m=\frac{\pi}{nL}(m+\frac{1}{2})-\frac{\mathrm{i}}{2nL}\ln\frac{n+1}{n-1}
\end{equation} 
where $m$ is an integer.  

Using the identity 
\begin{equation}
\frac{1-r^2}{1+r^2+2r\cos(x)}=\sum_{j=-\infty}^{\infty}(-r)^{|j|}
\mathrm{e}^{\mathrm{i} j x}
\end{equation}
valid for $|r|<1$ one can rewrite (\ref{dens1d}) as the trace formula for resonances in the considered simplest case
\begin{eqnarray}
&&\frac{1}{\pi}\sum_{m=\infty}^{\infty}
\frac{\mathrm{Im}k_m}{(k-\mathrm{Re}k_m)^2+ (\mathrm{Im}k_m)^2}\nonumber\\ 
&&=\frac{nL}{2\pi k}
\sum_{j=-\infty}^{\infty}(-r)^{|j|}\mathrm{e}^{2\mathrm{i} j nL k}\ . 
\label{simplest}
\end{eqnarray} 
The term with $j=0$ in this equation corresponds to the mean density of
resonances and terms with $j\neq 0$ describe the contributions from the
unique primitive periodic orbit and its repetitions.  

Notice that the transition from the Krein formula (\ref{dens1d}) to the
trace formula (\ref{simplest}) is done by removing from the
$\mathbf{S}$-matrix (\ref{matS1d}) the factor 
$ \mathbf{S}_0(E)=-\mathrm{e}^{-2\mathrm{i}kL}$ 
 which is exactly the $\mathbf{S}$-matrix for the scattering on the same
 cavity but with the Dirichlet boundary condition at the right end of the
 cavity $x=L$. This is the same phenomenon as for the circular dielectric
 cavity discussed in Section~\ref{weyl}.

\end{document}